\documentclass[14pt]{article} % Computer Modern font calls
\usepackage{amsmath,amssymb}
\usepackage{ifpdf}
\usepackage{hyperref}
\usepackage{graphicx}
\usepackage{mwe}
\usepackage{caption}

\newcommand\mysection{\setcounter{equation}{0}\section}

\def\baeq{\begin{appeq}}     \def\eaeq{\end{appeq}}
\def\baeeq{\begin{appeeq}}   \def\eaeeq{\end{appeeq}}
\newenvironment{appeq}{\beq}{\eeq}
\newenvironment{appeeq}{\beeq}{\eeeq}

\newcounter{Ahran}

\newcounter{hran}

\def\kps{\relax\ifmmode{{k_\perp^2}}\else{$k_\perp^2${ }}\fi}

\def\stot{\sigma_{\mbox{\scriptsize tot}}}

\def\ben{\begin{enumerate}}  \def\een{\end{enumerate}}
\def\bit{\begin{itemize}}    \def\eit{\end{itemize}}
\def\beq{\begin{equation}}   \def\eeq{\end{equation}}
\def\beql#1{\beq\label{#1}}    
\def\beeq{\begin{eqnarray}}  \def\eeeq{\end{eqnarray}}
\def\bq{\begin{quote}}       \def\eq{\end{quote}}

\def\eqref#1{(\ref{#1})}

\newskip\humongous \humongous=0pt plus 1000pt minus 1000pt
\def\caja{\mathsurround=0pt}

\newif\ifdtup

\def\eqal2#1{\,\vcenter{\openup1\jot
\caja   \ialign{\strut \hfil$\displaystyle{##}$&\hfil$
\displaystyle{{}##}$\hfil &$
\displaystyle{{}##}$\hfil\crcr#1\crcr}}\,}

\def\eV{{\rm e\kern-0.12em V}}  \def\GeV{{\rm G}\eV}

\def\be{\relax\ifmmode\beta\else{$\beta${ }}\fi}

\def\LQCD{\Lambda_{\mbox{\scriptsize QCD}}}

\def\cO#1{{\cal{O}}\!\left(#1\right)}

\def\al{\alpha}
\def\as{\al_s}

\def\half{{\textstyle \frac12}}

\def\ee{\relax\ifmmode{e^+e^-}\else{${e^+e^-}${ }}\fi}
\def\qq{\relax\ifmmode{q\overline{q}}\else{$q\overline{q}${ }}\fi}
\def\cR{{\cal{R}}}

\def\lrang#1{\left\langle #1 \right\rangle}
\def\kp{\relax\ifmmode{k_\perp}\else{$k_\perp${ }}\fi}

\def\ga{\mathrel{\mathchoice {\vcenter{\offinterlineskip\halign{\hfil
$\displaystyle##$\hfil\cr>\cr\sim\cr}}}
{\vcenter{\offinterlineskip\halign{\hfil$\textstyle##$\hfil\cr
>\cr\sim\cr}}}
{\vcenter{\offinterlineskip\halign{\hfil$\scriptstyle##$\hfil\cr
>\cr\sim\cr}}}
{\vcenter{\offinterlineskip\halign{\hfil$\scriptscriptstyle##$\hfil\cr
>\cr\sim\cr}}}}}

\def\lrang#1{\left\langle #1 \right\rangle}

\def\cZ{{\cal{Z}}}

\def\ib#1#2#3{{\em ibid.}~\underline{#1} (#3) #2}
\def\np#1#2#3{{\em Nucl.Phys.}~\underline{B#1} (#3) #2}
\def\pl#1#2#3{{\em Phys.Lett.}~\underline{#1B} (#3) #2}
\def\pr#1#2#3{{\em Phys.Rev.}~\underline{D#1} (#3) #2} 

\def\prep#1#2#3{{\em Phys.Rep.}~\underline{#1} (#3) #2}

\def\spj#1#2#3{{\em Sov.Phys.JETP}\/~\underline{#1} (#3) #2}

\def\zp#1#2#3{{\em Zeit.Phys.}~\underline{C#1} (#3) #2}

\def\epj#1#2#3{{\em Eur. Phys.J.}\/ {\underline {C#1}} (#3) #2}

 \def\cite#1{[\ref{#1}]}
 \def\citd#1#2{[\ref{#1},\ref{#2}]}
 \def\citt#1#2#3{[\ref{#1},\ref{#2},\ref{#3}]}
 
 \def\citm#1#2{[\ref{#1}--\ref{#2}]}

\newcounter{fcount}
\begin{document}
\begin{titlepage}

\begin{flushright}
``the full monty" \\
{\em the most or best that you can \\ have, do, get, or achieve, \\ or all that you want or need}\\[1 mm]
Cambridge Dictionary
\end{flushright}

\vspace*{\fill}

\begin{center}
{\Large\bf QCD-inspired description of multiplicity distributions in jets}
\end{center}
\par \vskip 5mm
\begin{center}
        {\bf\boldmath Yu.L.\ Dokshitzer$^*$}\\  
        Institute of Particle Physics and Accelerator Technologies,\\
        Riga Technical University, 7 Paula Valdena, Riga, Latvia
        \\
        \vskip 0.3 cm
        {\bf B.R.\ Webber}\\
        Cavendish Laboratory, University of Cambridge,\\
        J.J. Thomson Avenue, Cambridge, U.K.
        \end{center}
\par \vskip 6mm
\begin{center} {\large \bf Abstract} \end{center}
\begin{quote}
  We suggest a universal QCD-motivated expression for the Polyakov-KNO multiplicity distributions of hadrons in jets and compare it with data from \ee\ annihilation experiments.  The moments and overall shape of the distributions in full events and quark and gluon jets, over a range of energies, are described with reasonable quantitative precision. In particular, the scaling violation predicted by QCD is seen clearly in the moments and high-multiplicity fluctuations.
 \vfill
  
\end{quote}
\vspace*{\fill}

 \hrule
 \bigskip
 
$^*$ on leave from St Petersburg Nuclear Physics Institute,  Russia

\end{titlepage}

\tableofcontents

\mysection{Introduction}

In a recent publication \cite{DW25} we initiated a programme aimed at the description of multiplicity distributions in hard processes by means of perturbative QCD
% We have shown that a better treatment of the energy balance in the development of quark-gluon cascades allows one 
and demonstrated how to describe, in a parameter-free way, the pattern of {\em high-multiplicity fluctuations}.
We generalised the perturbative QCD analysis of multiplicity fluctuations in a gluon jet \cite{D93} to arbitrary ensembles of jets with a commensurate hardness scale. This includes a single quark jet, \qq\ pairs as the basis of \ee\ annihilation into hadrons, and gluon ensembles (from hadronic Higgs or $\Upsilon$ decays).
Single-quark and gluon jets manifest themselves in multiplicity distribution in one hemisphere of \ee annihilation and in high-$p_t$ LHC jet physics. 

In this paper, we extend the analysis to include
% of \cite{DW25}. This extension includes 
multiplicity moments as well as the entire shape of the corresponding multiplicity distributions. 
 
In Section \ref{Sec:PKNO} we remind the reader of the scaling phenomenon in the structure of multiplicity distributions known as KNO scaling \cite{KNO} and discuss its status in the domain of hard interactions.

In Section \ref{Sec:MDLA} the perturbative QCD framework is set up which we refer to as MDLA --- the modified double-logarithmic approximation. As shown in \cite{DW25}, MDLA provides a high-fidelity description of the tail of multiplicity distributions.

Here we confront with experiment the MDLA prediction for mean powers of charged particle multiplicity ---multiplicity moments --- and suggest a compact QCD-motivated formula that describes not only the high-multiplicity tail but the entire shape of multiplicity distributions in hard processes in a universal way.

Section \ref{Sec:Exp} is devoted to comparisons of these distributions with the data on multiplicity fluctuations in \ee\ annihilation as well as single-quark and gluon jets.

\smallskip

Our aim is to demonstrate that the QCD dynamics of parton multiplication is in reasonable quantitative agreement with observation.

\mysection{The P-KNO phenomenon \label{Sec:PKNO}}
\def\bn{\bar{n}}
Various high-energy interactions give rise to the production of a large number of particles. 
The basic quantity that characterises multiparticle production is the probability distribution
\beq\label{eq:Pn}
   P_n(Q) \equiv \frac{\sigma_n(Q)}{\stot(Q)} 
\eeq
with $Q$ the energy scale of the process. 
When the number of final-state particles is large, one can approximate the discrete distribution \eqref{eq:Pn} by a continuous function of the ratio of the number of produced particles to the mean multiplicity,
\beq
 \nu \equiv \frac{n}{N(Q)} , \quad N(Q) = \sum_{n=1}^\infty nP_n(Q)  \>\equiv \lrang{n}\!(Q) ; \qquad 
 \sum_{n=0}^\infty P_n \cdots \>\Longrightarrow\> \int_0^\infty d\nu\,\big[ N(Q)\,P_n(Q)\big] \cdots
\eeq
What we know as the KNO scaling~\cite{KNO} is the hypothesis that the entire dependence of the probability distribution on the collision energy is absorbed by the energy dependence of the mean multiplicity in the form
\beq\label{eq:Psidef}
   N(Q)\cdot P_n(Q) \>\equiv \> \Psi\big(\nu,\, Q\big) \>\stackrel{\mbox{\scriptsize KNO}} {\to} \>  \Psi\big(\nu\big) .
\eeq
In the field of hard interactions, the scaling law \eqref{eq:Psidef} was predicted by A.M.~Polyakov, who considered \ee\ annihilation into hadrons in the framework of an abstract conformal QFT back in 1970 \cite{AMPolyakov}.
In order to separate the hard and soft interaction domains where the physics of particle production is essentially different, we proposed to refer to the law \eqref{eq:Psidef} in hard processes as P-KNO scaling.

With the advent of QCD, A. Bassetto, M. Ciafaloni, and G. Marchesini came to the conclusion \cite{BCM} that P-KNO scaling 
has to hold asymptotically as a consequence of the cascading nature of parton multiplication. 
An analytic solution was obtained later in \cite{DFK} in the leading double-logarithmic approximation (DLA) of perturbative QCD (see \cite{Book} for a review).

For hard interactions, $Q$ marks the hardness of the process: the characteristic momentum transfer rather than the energy. In the case of \ee\ annihilation the two coincide: $s=Q^2$. In a more general case, $Q$ measures the maximal transverse momentum available for parton cascading. In particular, for a jet with energy $E$ and opening angle $\Theta$ one has
\beq\label{eq:hardness}
  Q = 2E\sin\frac\Theta{2}.
\eeq
For small $\Theta$ this expression reduces to the maximal transverse momentum of a parton belonging to the jet, $k_\perp\simeq k\theta$, $k\le E$, $\theta\le\Theta$. For \ee\ annihilation viewed as a \qq\ pair as a source of multi-hadron production, one sets $\Theta=\pi$ to obtain $Q=2E=\sqrt{s}$. If one seeks to isolate a single quark jet by looking in one hemisphere, then $\Theta=\pi/2$, yielding $Q=\sqrt{s/2}$.

\subsection{Multiplicity moments and Laplace transform}

A convenient tool for counting multiplicities is the generation function
\beq\label{eq:GF}
  \cZ(u,Q) = \sum_{n=0}^\infty P_n(Q)\, u^n \>\simeq\> \int_0^\infty d\nu\, \bigg[ N(Q)\, P_n(Q)\bigg]\cdot u^{N(Q)\cdot\nu } \>\equiv\> \int_0^\infty d\nu\, \Psi(\nu,Q)\cdot u^{N(Q)\cdot\nu }.
\eeq
By repeated differentiation with respect to $u$ at $u=1$, one obtains the {\em factorial moments} of the multiplicity distribution:
\beq\label{eq:facmom}
\left.\left(\frac d{du}\right)^k\cZ(u,Q)\right|_{u=1} 
= \lrang{n(n-1)\cdots(n-k+1)}(Q)\,.
\eeq
On the other hand, by repeated differentiation with respect to $\ln u$, one obtains the {\em power moments}:
\beq\label{eq:powmom}
\left.\left(u\frac d{du}\right)^k\cZ(u,Q)\right|_{u=1} 
= \lrang{n^k}(Q)\,.
\eeq
Writing $\beta=-N(Q)\ln u$, we see that the generating function, considered as a function of $\beta$, is just the Laplace transform of the P-KNO distribution:
\beq\label{eq:subs}
 \cZ\left(u=\exp\left\{\frac{-\beta}{N(Q)}\right\},Q\right) \>=\> \int_0^\infty d\nu\, \Psi(\nu,Q)\, e^{-\beta\cdot\nu }\equiv\Phi(\beta,Q)\,.
\eeq
The P-KNO power moments are then obtained by differentiation w.r.t. $\beta$ at $\beta=0$:
\beq\label{eq:KNOmom}
\left.\left(-\frac d{d\beta}\right)^k\Phi(\beta,Q)\right|_{\beta=0}
=\frac{\lrang{n^k}}{[N(Q)]^k}\equiv g_k(Q)\,,
\eeq
and the generating function has a Taylor expansion in terms of the P-KNO power moments
\beq\label{eq:phiexp}
\Phi(\beta,Q) \>=\> \sum_{k=0}^\infty \frac{(-\beta)^k}{k!}\cdot g_k(Q)\,.
\eeq
We conclude that strict KNO scaling holds if and only if the l.h.s.\ of this equation, that is the generating function itself, does not depend on energy (or hardness scale for P-KNO).\footnote{Modulo, of course, the possibility of replacing a discrete sum by an integral, for which substitution $n$ should be numerically large.}
Usually we expect such scaling to hold in the high-energy limit, but formally speaking, this is not even necessary.  In fact, we shall see that a well-defined pattern of scaling violation is expected at sub-asymptotic energies.

\subsection{DLA multiplicity moments}

The exact asymptotic values of the moments $g_k$ can be obtained successively using the recurrence relations that follow from the DLA solution of the P-KNO problem.

For the multiplicity moments of high rank $k$, a quite simple asymptotic formula applies~\citd{DFK}{Book}.
In particular, for a gluon jet 
\begin{subequations}\label{eq:DLAgk}
\beq\label{eq:gkgdla}
   g_k = \frac{2\,k!}{C^k}\left(k+ \frac1{3k}\right)\cdot\left[ 1 + \cO{k^{-4}} \right] , \quad C\approx 2.552 .
\eeq
For the case of a particle source $S$ other than a gluon, the following % general 
relation holds:
\beq\label{eq:Phirho}
\Phi^{(S)}(\beta) \>=\> \left[ \Phi\left(\frac\beta{\rho_S}\right) \right]^{\rho_S} , \qquad 
g_k^{(S)} = \left. \frac{d^k}{d\beta^k} \left[ \Phi\left(\frac{-\beta}{\rho_S}\right) \right]^{\rho_S} \right|_{\beta=0}.
\eeq
\end{subequations}
Here $\rho_S$ is the ratio of mean particle multiplicities,
\beq\label{eqrhodef}
  \rho_S = \lrang{n_S}(Q)\big/\lrang{n_g}(Q) , \qquad S = q,\> \qq, \> gg,\> \mbox{\em etc}.
\eeq
Hereafter we refer to the parameter $\rho_S$ as the {\em hadron production power} of the source $S$.

For the quark, we use $\rho_q=2/3$, a value that is close to the quark-to-gluon ratio of mean charged-hadron multiplicities at the hardness scale $Q\sim 100\,\GeV$. 

Surprisingly, for all relevant particle sources, the asymptotic formulae \eqref{eq:DLAgk} were found to reproduce the exact low-rank multiplicity moments quite well starting from the lowest rank, $k\ge1$ (a comparison can be found in Appendix B of \cite{DW25}).

Low-rank moments are practically interesting in that they quantify the shape of the P-KNO distribution by means of dispersion ($k=2$), skewness ($k=3$), kurtosis ($k=4$), etc.
On its own merits, an observation of the precocious asymptotics has only academic interest. The DLA is well-known to be fatally inappropriate for comparison with observation. 
The DLA corresponds to the limit $\as\to 0$, whereas the effects of the finite QCD coupling are crucial. 
The standard order-by-order evaluation of perturbative corrections in $\sqrt\as$ --- the characteristic expansion parameter for the quantities dominated by soft gluon radiation --- converges badly \cite{MW84} and does not offer a reliable approximation. 

\def\DLA{\mbox{\scriptsize DLA}}
\def\MDLA{\mbox{\scriptsize MDLA}}

\mysection{MDLA approach to P-KNO \label{Sec:MDLA}}

An alternative approach to the problem was suggested in \cite{D93}. It consists of an all-order resummation of the terms of the perturbative QCD series that are proportional to the rank of the multiplicity moment considered, $(\sqrt\as \cdot k)^m$. 
The modified double logarithmic approximation (MDLA) takes care of the energy balance in parton multiplication that is systematically ignored in DLA.

The MDLA predictions depend on the hardness scale $Q$ of the process via the anomalous dimension of the mean particle multiplicity 
\beq\label{eq:gammadef}
  \gamma(\as(Q)) \>\equiv\> \frac{d \ln N(Q)}{d\ln Q} .
\eeq
In particular, gluon jet multiplicity moments acquire in this approximation a factor % \cite{D93}
\beq\label{eq:gkgmdla}
g^{(g)}_{k,\MDLA} = g^{(g)}_{k,\DLA} \cdot \frac{[\Gamma(1+\gamma)]^k}{\Gamma(1+\gamma k)}  
\eeq
with the DLA moments given in \eqref{eq:gkgdla}.  The MDLA generating function is obtained by the corresponding substitution in Eq.~\eqref{eq:phiexp}, and the moments for a general source $S$ are obtained from the relation \eqref{eq:Phirho}.\footnote{Strictly speaking, this relation has only been established at DLA. The question of how much this assumption might affect the MDLA results for $\rho\neq 1$ is presently under study.}

\subsection{Low-rank multiplicity moments \label{Sec:LowMoms} }
Given the unexpected precision of the asymptotic expression for low-rank DLA multiplicity moments, we decided to look into the effect that the MDLA modification brings into the game. The results were quite surprising. 
Not only do the MDLA factors introduced in \eqref{eq:gkgmdla} bring the low-rank multiplicity moments closer to reality, which was expected, but the first few moments are quite close to the observed values. % $k=2,\ldots 5,6$.

This is shown in Fig.~\ref{fig:momsOPAL}, where the first five normalized power moments measured by LEP-1 collaborations \citt{OPAL1992}{DELPHI1}{L3} are compared with the MDLA and DLA predictions.\footnote{The
L3 power moments in Fig.~\ref{fig:momsOPAL} are recalculated from the values of the {\em factorial}\/ moments (up to rank $k=6$) presented in \cite{L3}.}

 Fig.~\ref{fig:momsOPAL1} shows an analogous comparison for fluctuations in a single hemisphere.

\medskip

\noindent
\begin{minipage}{0.5\textwidth}
\begin{center}
\includegraphics[width=\textwidth]{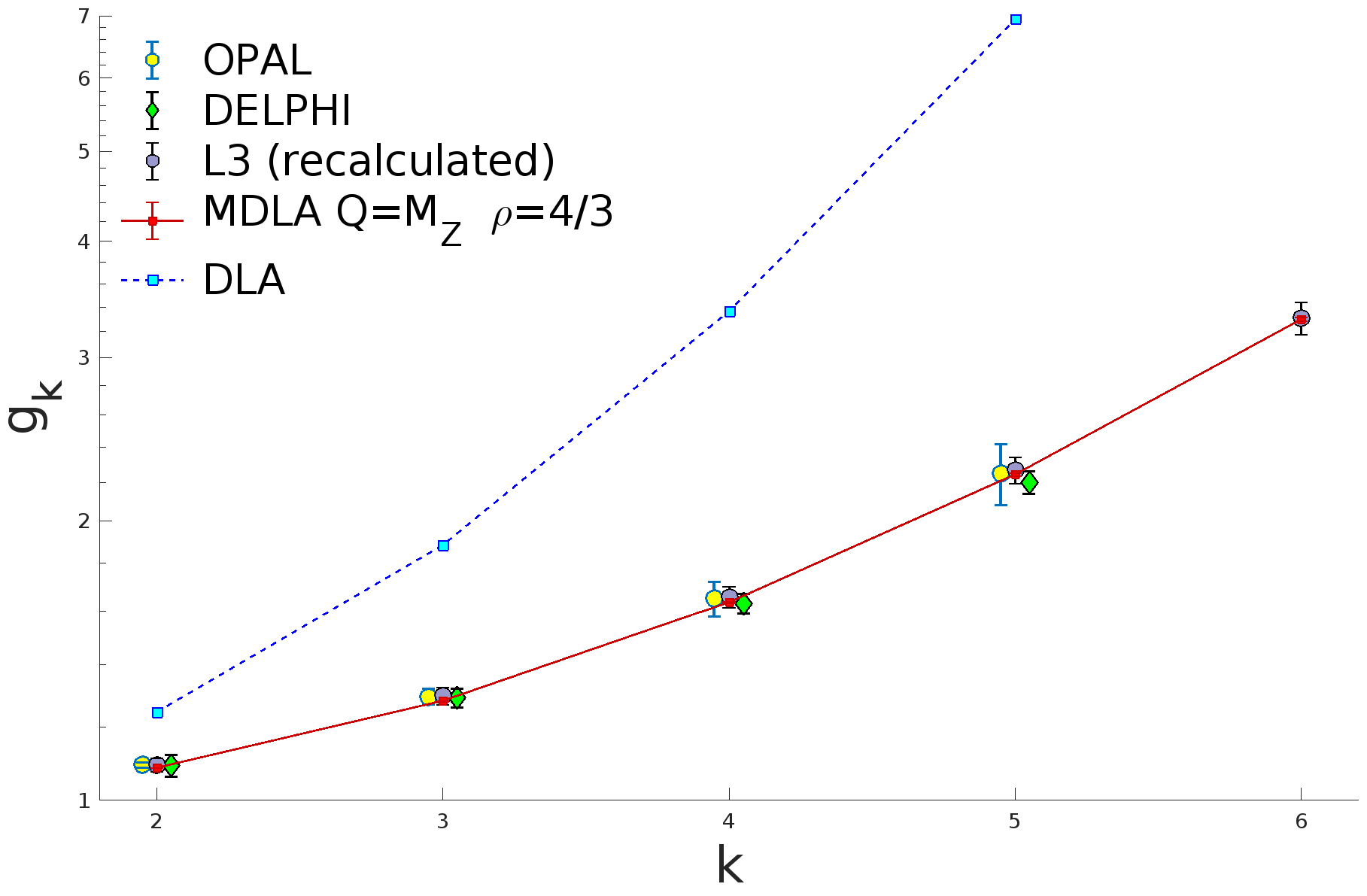}
\captionof{figure}{LEP-1 multiplicity moments \label{fig:momsOPAL}}
\end{center}
\end{minipage}
\begin{minipage}{0.5\textwidth}
\begin{center}
\includegraphics[width=\textwidth]{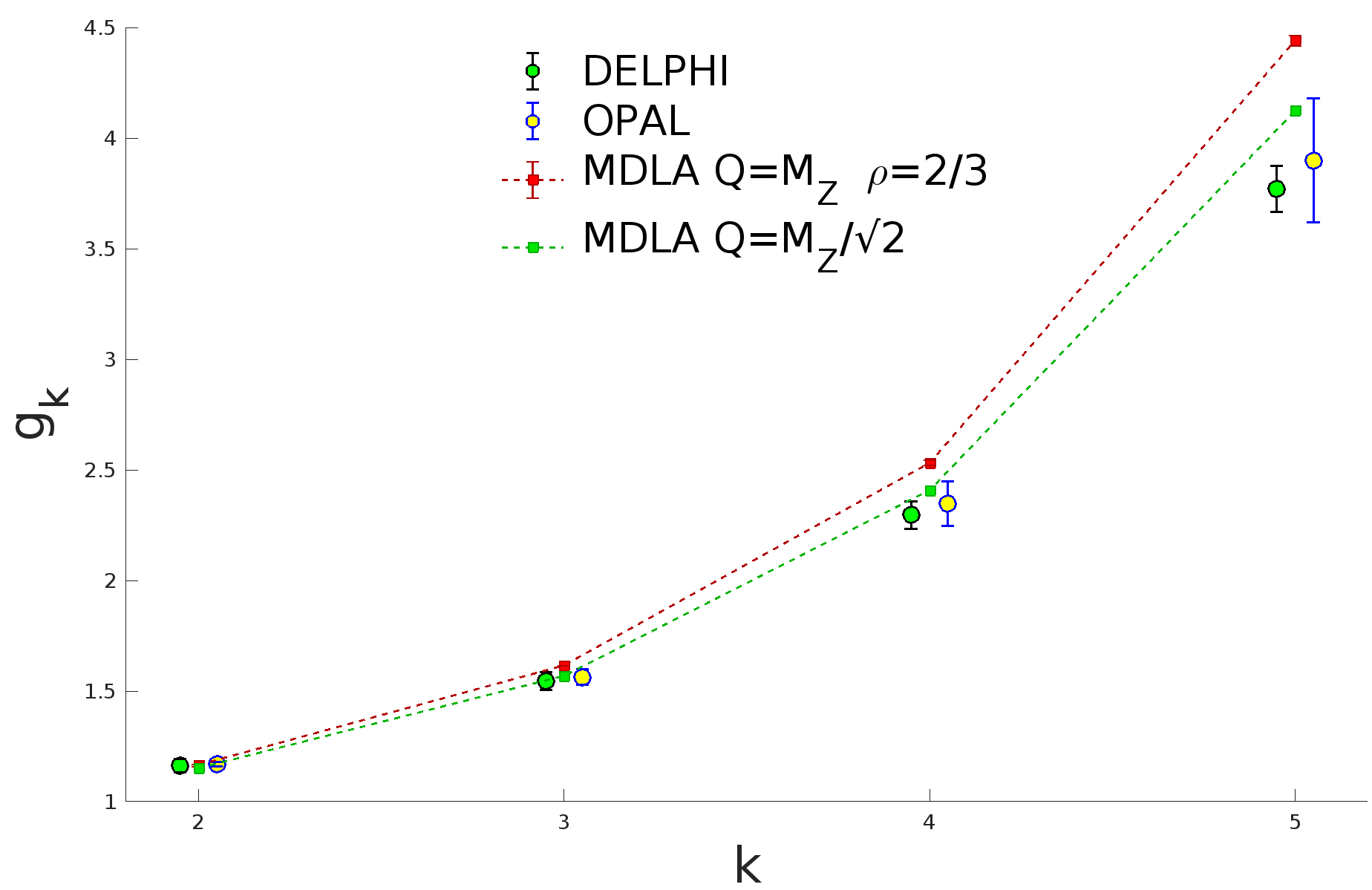}
\captionof{figure}{LEP-1 one-hemisphere moments\label{fig:momsOPAL1}}
\end{center}
\end{minipage}
\medskip

In Fig.~\ref{fig:momsOPAL1} the theoretical prediction with a reduced hardness scale $Q\to Q/\sqrt2$ is also shown.
Such a scale reduction was argued in \cite{DW25} to be appropriate for a theoretical approach to the description of high-rank multiplicity moments of well-developed parton cascades. Whether that argument can be applied to low ranks is unclear.
The $\sqrt2$ rescaling factor does not affect the result much, but in the right direction.
In any case, the difference lies beyond the most optimistic expectation of the potential accuracy of MDLA considerations.

The first six moments measured at a lower scale, $Q=29\,\GeV$, by the HRS Collaboration~\cite{HRSmoms} at SLAC are shown in Fig.~\ref{fig:momsHRS}.
\begin{center}
\includegraphics[width=0.7\textwidth]{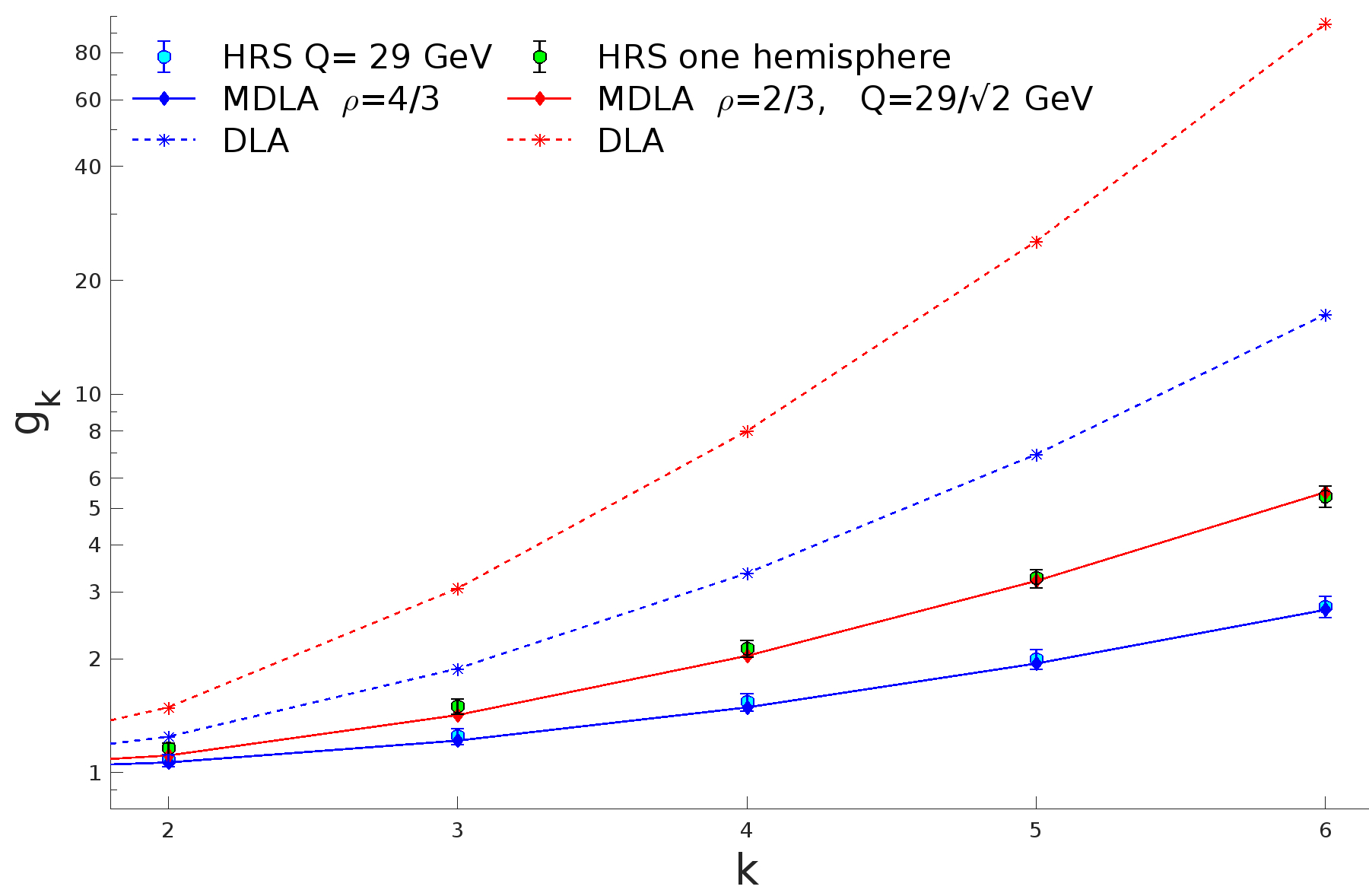}
    \vspace{-3mm}
\captionof{figure}{\label{fig:momsHRS}HRS full event and one-hemisphere multiplicity moments \cite{HRSmoms}}
\end{center}
%%%
We conclude that the MDLA significantly improves the QCD description of low-rank moments.
\medskip

One can also look at a commonly used measure of the shape of the multiplicity distributions, which demonstrates how drastically different they are from the asymptotic QCD regime. This is the ratio of mean multiplicity to the dispersion:
\beq
N/D = \lrang{n}/\sqrt{\lrang{n^2}-\lrang{n}^2} = 1/\sqrt{g_2-1}\,.
\eeq
This observable provides a most brutal challenge for the theory. As may be seen in Fig.~\ref{fig:momsHRS}, $g_2$ at low scales is close to unity, and in fact the MDLA corrections drive $g_2-1$ negative for $\gamma>0.56$ ($Q\lesssim 5$ GeV).

The problem is illustrated by Fig.~\ref{fig:ND} where experimental measurements of the $N/D$ ratio are presented together with theoretical predictions of DLA and MDLA. 
\begin{center}
\includegraphics[width=0.75\textwidth]{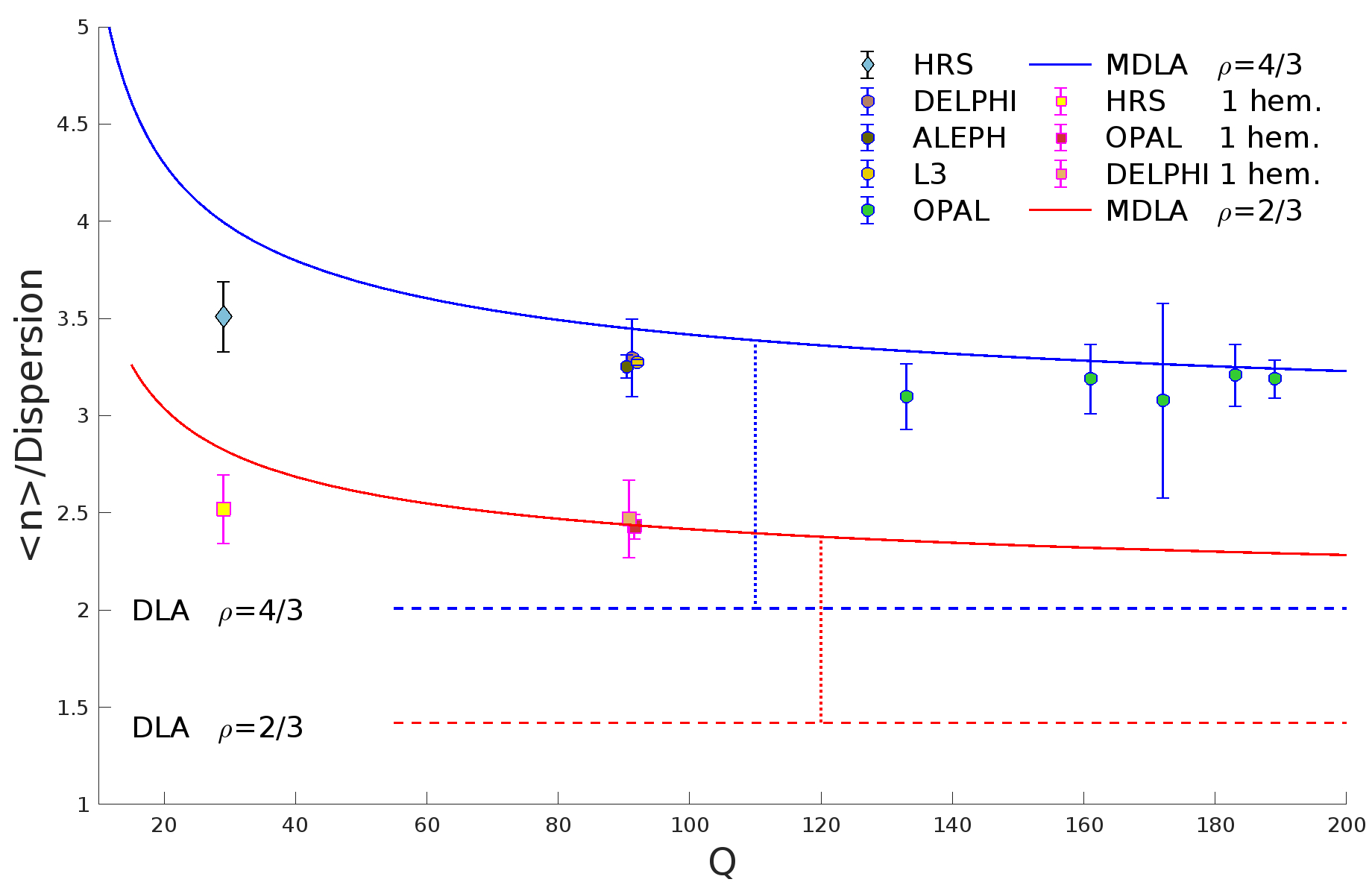}
    \vspace{-3mm}
\captionof{figure}{\label{fig:ND}$<\!N\!>/D$ ratio } % \cite{HRSmoms}}
\end{center}
This busy picture needs explanation. 
Horizontal dashed lines mark the scale-independent asymptotic DLA values for the relevant production power parameters $\rho= 2/3, 4/3$. 
Solid lines show the MDLA predictions. 
The gaps between the DLA and MDLA for the full event ($=\qq$) and for one hemisphere ($S=q$) are marked by dotted lines to guide the eye.

On the positive side, one may conclude that the introduction of the MDLA correction factor makes the discrepancy with experiment disappear at sufficiently large hardness scales. Meanwhile, it implies a strong violation of KNO scaling below $Q=\cO{30 - 40\,\GeV}$ driven by proximity to the critical point $\gamma\simeq 0.56$. 
Both HRS points (29 \GeV) fall below the corresponding MDLA curves but, at the same time, may indicate a certain scaling violation.

The false singularity makes MDLA predictions for moderate hardness scales extremely sensitive to the precise functional dependence of the anomalous dimension on $\as$ and to the momentum dependence of the QCD coupling itself. 

It is clear that a more refined analysis is needed to address the physics of relatively low scales.

\subsection{Shape of the P-KNO distributions \label{Sec:Shape}}

\subsubsection{ The tail $\nu > 1$\label{Sec:Tail}}
The perturbative expansion of the MDLA factor in \eqref{eq:gkgmdla} resums powers of the product $(k\gamma)$. At the same time, this approximation ignores non-enhanced $(\sqrt\as)^n$ terms. This makes MDLA suitable for studying high-multiplicity fluctuations that correspond to large moment ranks $k\gg1$. 

\def\ksd{k_{\mbox{\scriptsize sd}}}

\medskip

The probability of multiplicity fluctuations in a gluon jet,
\begin{subequations}\label{eq:PolExp}
    \beeq\label{eq:kappadef}
\nu\Psi(\nu,\gamma) &\!\! \propto  \!\! &  e^{- \kappa(\nu,\gamma)}, \qquad \kappa = \big[ D\nu\big]^\mu, \\
\ D & \!\! = \!\! & D(\gamma)=\frac{C\,\gamma^\gamma(1-\gamma)^{1-\gamma}  }{ \Gamma(1+\gamma) }  ,\qquad \mu\>=\> \mu(\gamma) = \frac1{1-\gamma},
\eeeq
\end{subequations}
falls faster than exponentially, with the power of $\nu$ in the exponent of \eqref{eq:kappadef} depending on the multiplicity anomalous dimension in the way predicted by A. Polyakov~\citd{AMPolyakov}{D93}.

The approximate analytic expression for the P-KNO tail reads (see App.~D of \cite{DW25} for details) 
\begin{subequations}\label{eq:tail}
\beq\label{eq:Psi}
\nu\Psi(\nu) \>=\> 2\mu^2e^{-\kappa} \chi(\ksd) \cdot (\kappa^2-\kappa) \cdot \bigg[ 1+ \cO{\kappa^{-2}} \bigg].
\eeq
Here the factor $\chi(k)$, slowly changing with the moment rank $k$, is given by
\beq\label{eq:chidef}
  \chi(k,\gamma) = \frac{\Gamma(1+k)[\gamma^\gamma(1-\gamma)^{1-\gamma}]^k}{  \Gamma(1+\gamma k) \Gamma(1+(1-\gamma)k)} ,
\eeq
and $\ksd$ is the characteristic moment rank that emerges from the steepest descent evaluation of Laplace moments of the distribution that have to match \eqref{eq:gkgmdla} at large ranks:
\beq\label{eq:ksd}
\ksd = \mu\,\big(\kappa-2\big)  \cdot \bigg[ 1+ \cO{\kappa^{-2}} \bigg].
\eeq
\end{subequations}
It is important to stress that the scale dependence of the perturbative QCD P-KNO distribution is contained in a single quantity, that is, the multiplicity anomalous dimension $\gamma(\as(Q))$.  
The dependence on $\nu$ is embedded in the characteristic combination
\beq\label{eq:kappadef2}
    \kappa \equiv \big[ D(\gamma)\,\nu\big]^{\mu(\gamma)}.
\eeq
The tail formula \eqref{eq:tail} is applicable to $\nu \ga 1$. 
As shown in \cite{DW25}, MDLA is capable of describing the {\rm \bf right} slope of all P-KNO distributions measured at \ee accelerators in the past, as well as LHC jets.

\subsubsection{ Generalisation to entire range $0<\nu<\infty$ \label{Sec:DistrG}}
There is no solid approach to address the shape of the {\rm\bf left} slope of the P-KNO distribution, $\nu<1$.
The only theoretical result concerning the behaviour of $\Psi(\nu)$ for $\nu\to 0$ is the DLA asymptotic regime
\beq
  \Psi(\nu) \>\propto\> \exp\left\{ -\half \ln^2\nu \right\}.
\eeq
This prediction is purely academic.
It implies the limit $\lrang{n}(Q)=\infty$, which is far from realistic. 
For practical purposes, one can consider the $\nu<1$ interval as a playground where almost anything goes.

The expression \eqref{eq:tail} could not be used to fit the entire distribution in $\nu$ due to two problems: as it stands, neither $\Psi$ \eqref{eq:Psi} nor $\ksd$ \eqref{eq:ksd} is positively definite. 
In \cite{DW25} it was argued that dropping the subleading subtraction term ``--2" in $\ksd$ does not much affect the right slope of the P-KNO distribution. 
Still, there is an alternative: a simple generalisation that preserves the accuracy of the description of large moment ranks while enforcing positivity, which allows one to peek into the $\nu<1$ domain. 
It amounts to replacing 
\begin{subequations}\label{eq:p0p1}
\beeq\label{eq:ksdMod}
\ksd & \Longrightarrow & \mu \,\kappa^{1+p_0}\left[\kappa+\frac{2}{p_0}\right]^{-p_0} = \mu\,\big(\kappa-2\big)  \left[ 1+ \frac{2(p_0+1)}{p_0\kappa^2} +\ldots\right], \\
\label{eq:PsiMod}
(\kappa^2-\kappa) & \Longrightarrow &  \kappa^{2+p_1} \left[ \kappa +\frac{1}{p_1}\right]^{-p_1} = (\kappa^2-\kappa)\left[ 1+ \frac{(p_1+1)}{2p_1\kappa^2} +\ldots\right].
\eeeq
\end{subequations}
Varying $p_0$ and $p_1$ does not affect the high-multiplicity tail. At the same time, one can choose the values of these parameters to reproduce the shape of the {\em left slope}. More importantly, adjusting the two parameters, one can attempt to rescue the lowest moments $g_0=1$ (normalization) and $g_1=1$ (mean multiplicity sum rule).

The P-KNO distribution applicable to the full range of multiplicity fluctuations in a gluon jet then takes the form
\begin{subequations}\label{eq:PsiGS}
\beq\label{eq:PsiG}
\nu\Psi(\nu)\simeq \frac{2\mu^2  \,\kappa^{2}e^{-\kappa} }{\left[ 1 +(p_1 \kappa)^{-1}\right]^{p_1}}\, \chi(\ksd), \quad 
\ksd= \frac{ \mu\,\kappa}{\left[ 1 +2(p_0 \kappa)^{-1}\right]^{p_0}}.
\eeq

\subsubsection{P-KNO distribution for a general source \label{Sec:DistrRho}}

Once the gluon distribution is known, the P-KNO distribution for the case of a general source can be derived following the route described in Appendix E of \cite{DW25}.

For an arbitrary source $S$ with hadron production power $\rho$ one obtains
\beq\label{eq:PsiS}
\nu\Psi^{(S)}(\nu)\simeq  \big[\nu\Psi(\nu) \big]^{\rho}\,\cdot \sqrt{\rho} \, G^{\rho-1}, \quad 
G =\sqrt{ \frac{ 2 \pi \mu^{-2}}{ \gamma\kappa +2(1-\gamma)}}
\eeq
\end{subequations}

\mysection{Confronting the data \label{Sec:Exp}}

\subsection{The model \label{sec:model}}

\subsubsection{Fixing parameters}

In principle, Eq.\eqref{eq:PsiG} contains two arbitrary parameters $p_0$ and $p_1$ that are not theoretically controlled at our current level of ignorance about small-multiplicity fluctuations. 
It gives a certain freedom for attempts to fit the data in an optimal way.
This not being the purpose of the present study, we take a conservative approach. We fix the values 
\beql{eq:params}
    p_0=1,\quad  p_1=4 
\eeq
as source- and energy-independent parameters. In the following, the abbreviation MDLA refers to \eqref{eq:PsiGS} with the above parameter prescription \eqref{eq:params}.

Moreover, we ascribe to a quark the hadron production power $\rho_q=2/3$. It is not a sacred number. 
The ratio of gluon-to-quark mean charged-hadron multiplicities depends on the hardness scale of the process. 
The value $\lrang{n_g}/\lrang{n_q}\simeq 1.5$ belongs to the LEP hardness scale of around 100 \GeV\ where the bulk of the precision data came from. 

The multiplicity anomalous dimension $\gamma(\as)$ is given by the MLLA-expression \citm{AHM}{EAO}.
Quantitative details of this and all the other ingredients of the MDLA are given in the Appendices.

\subsubsection{Verification of the low moments }

In Section \ref{Sec:LowMoms} we compared analytic MDLA predictions for low-rank multiplicity moments with experiment. 
Now we have the expression \eqref{eq:PsiGS} that is supposed to describe the entire P-KNO distributions.
In the derivation of \eqref{eq:PsiGS} certain approximate intermediate steps were involved (namely, steepest descent and stationary phase evaluation of the inverse Laplace transform),
as well as introduction of the model parameters $p_{0,1}$.
Therefore, it is worth verifying whether the moments that one obtains by numerical integration of the distribution agree with the analytic predictions presented in Sect.~\ref{Sec:LowMoms}. 
\begin{center}
    \includegraphics[width=0.9\linewidth]{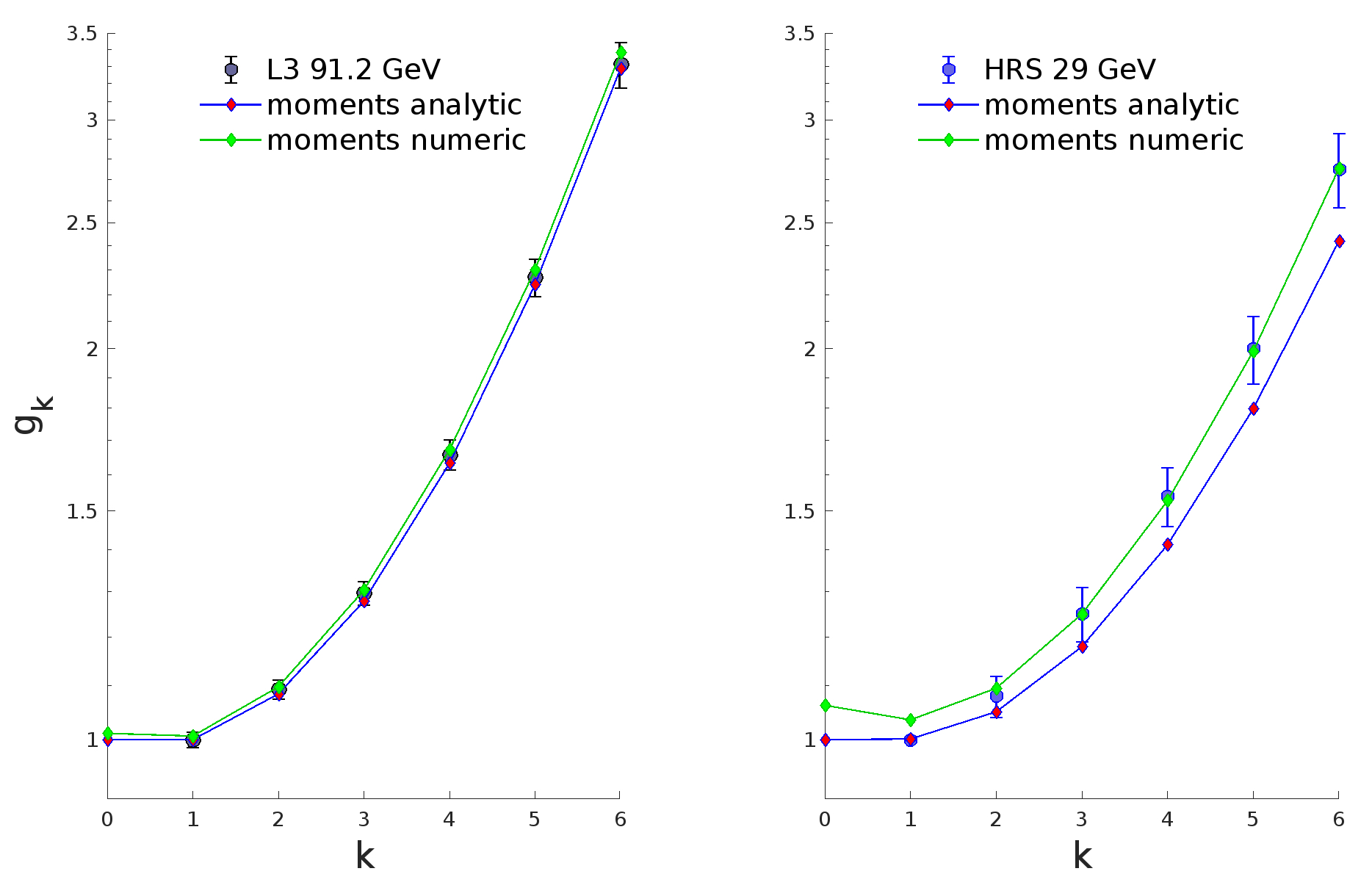}
    \captionof{figure}{Low-rank moments; numerical vs. analytic for $Q=91$ and $29\,\GeV$    \label{fig:MOMS_L3_HRS} }
 \end{center}
As shown in Fig.~\ref{fig:MOMS_L3_HRS}, the difference between numerical and analytic values, hardly visible at LEP energy, reaches a few percent for HRS (29 GeV). It is sensitive to the {\em ad hoc} subleading (in $1/\kappa$) terms in Eqs.\eqref{eq:p0p1}. 
The good agreement between the numerical moments $k\ge 2$ and the HRS data does depend on our choice of the parameters $p_0$ and $p_1$.

\subsection{\ee}

\subsubsection{$Q=M_Z$}
High-quality measurements of the P-KNO distribution in \ee\ annihilation were performed at the $Z^0$ peak. As shown in Fig.~\ref{fig:LEP1}, the results reported by all four LEP Collaborations \citm{ALEPHdist}{L3} agree.
\begin{center}
    \includegraphics[width=0.7\linewidth]{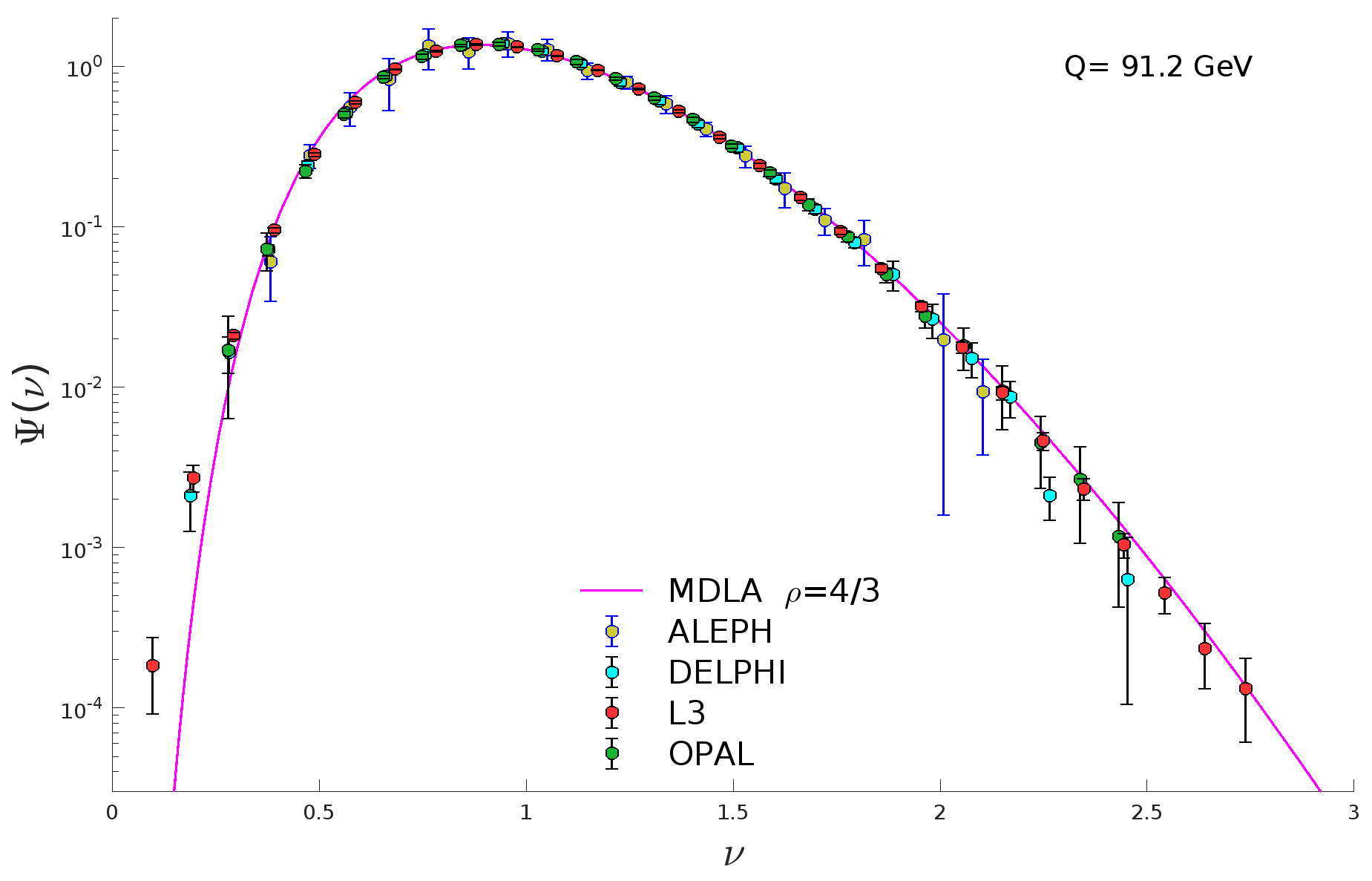}
    \captionof{figure}{P-KNO distribution at $\sqrt{s}=M_Z$ \label{fig:LEP1}   }     
\end{center}

In Fig.~\ref{fig:L3_KNO} we show the distribution with the smallest quoted errors,  obtained by L3 \cite{L3}, on both logarithmic and linear scales. The MDLA equation \eqref{eq:PsiGS} with $\rho=2\rho_q=4/3$ fits the data unexpectedly well throughout the entire range of charged hadron multiplicities.
\begin{center}
    \centering
    \includegraphics[width=0.9\linewidth]{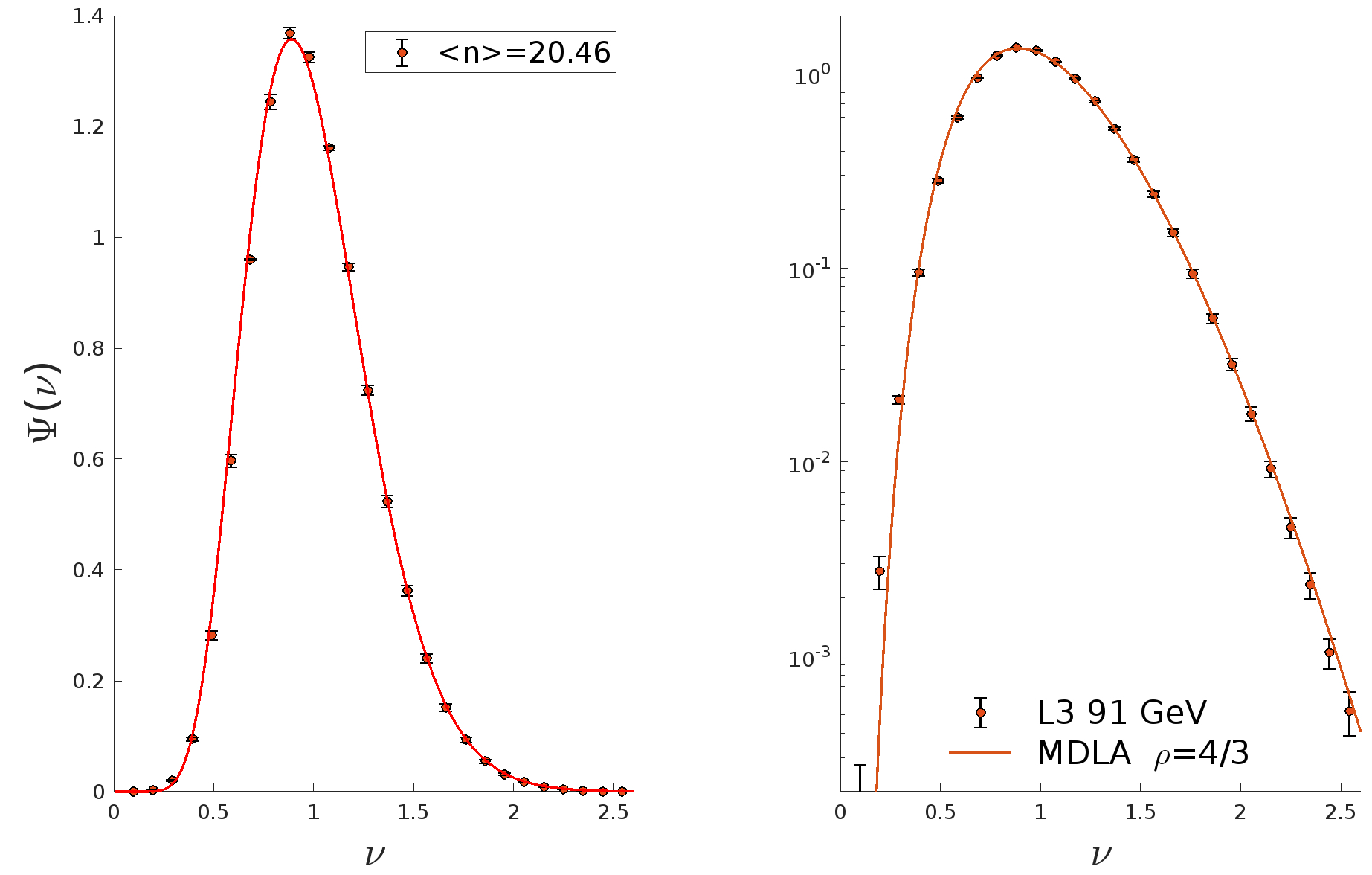}
    \captionof{figure}{ \label{fig:L3_KNO}  L3 P-KNO distribution of charged hadrons in \ee\ annihilation at the $Z$-peak}
\end{center}

\subsubsection{LEP-2}

Increasing the energy does not pose a problem for Eq.~\eqref{eq:PsiGS}; see Figs.~\ref{fig:OPAL_161}, \ref{fig:OPAL_189} where the OPAL data \cite{OPAL_LEP2} for $Q= 161, 189 \,\GeV$ are confronted with the theoretical prediction.
\begin{center}
    \includegraphics[width=0.7\linewidth]{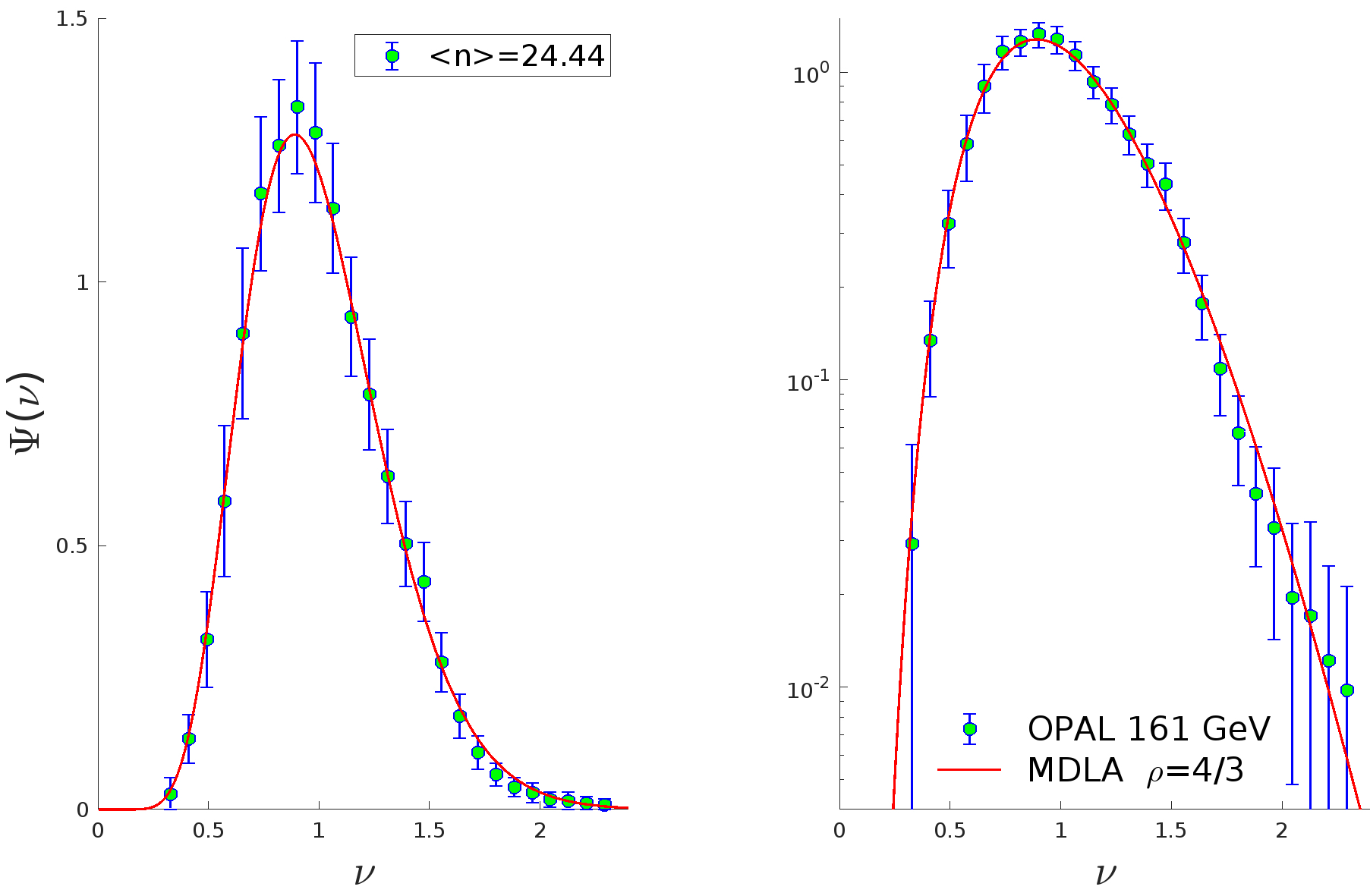}
    \captionof{figure}{P-KNO distribution at $\sqrt{s}=161\,\GeV$     \label{fig:OPAL_161} }
\end{center}
\begin{center}
    \includegraphics[width=0.8\linewidth]{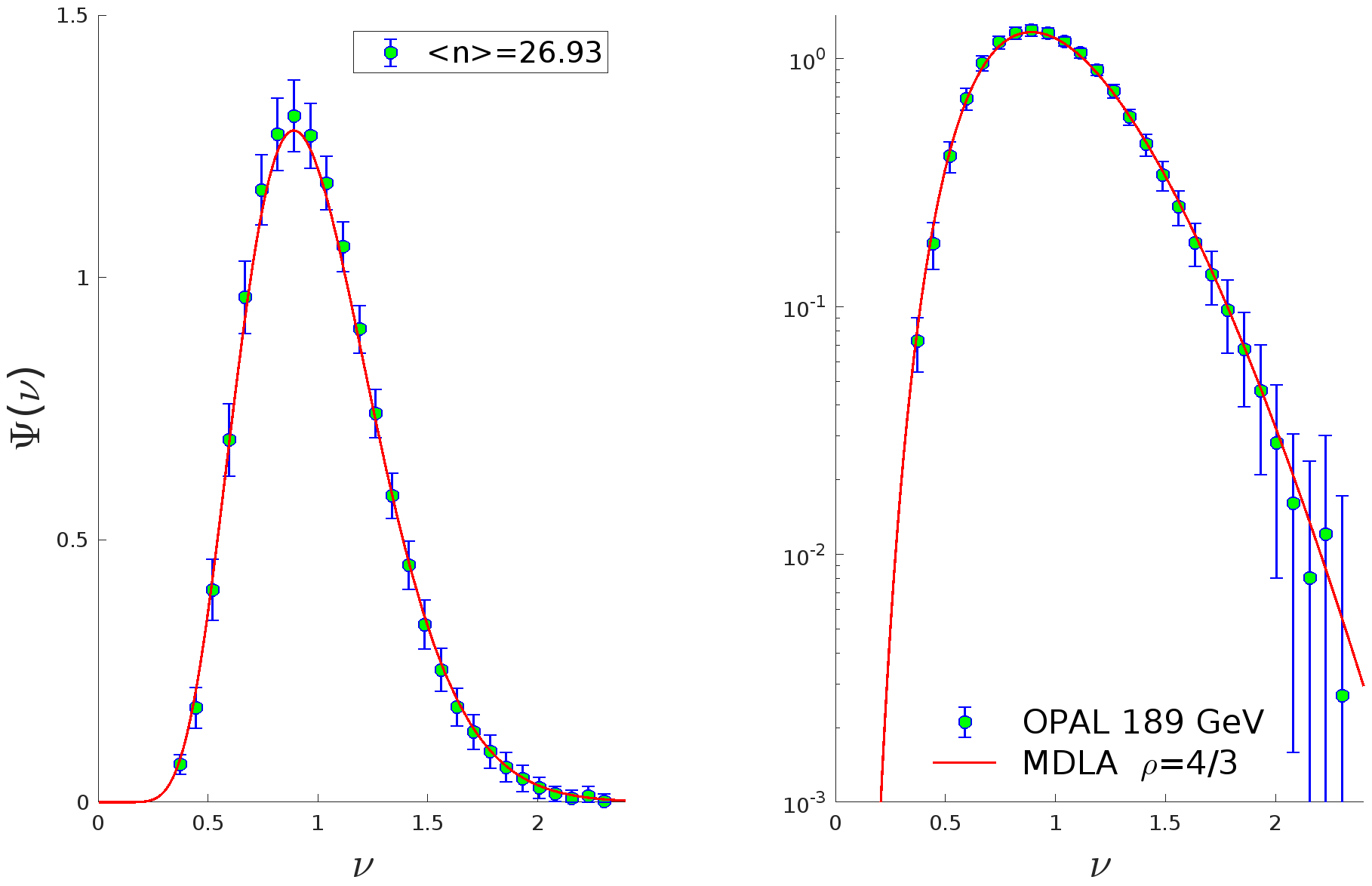}
   %     \vspace{-2mm}
    \captionof{figure}{P-KNO distribution at $\sqrt{s}=189\,\GeV$     \label{fig:OPAL_189} }
\end{center}

\subsubsection{Smaller hardness scales}

Similar comparisons for lower annihilation energies (TASSO, $Q=44\,\GeV$ and HRS, $Q=29\,\GeV$) are presented in Figs.~\ref{fig:TASSO_44} and Fig.~\ref{fig:HRS_29}.
\begin{center}
    \includegraphics[width=0.8\linewidth]{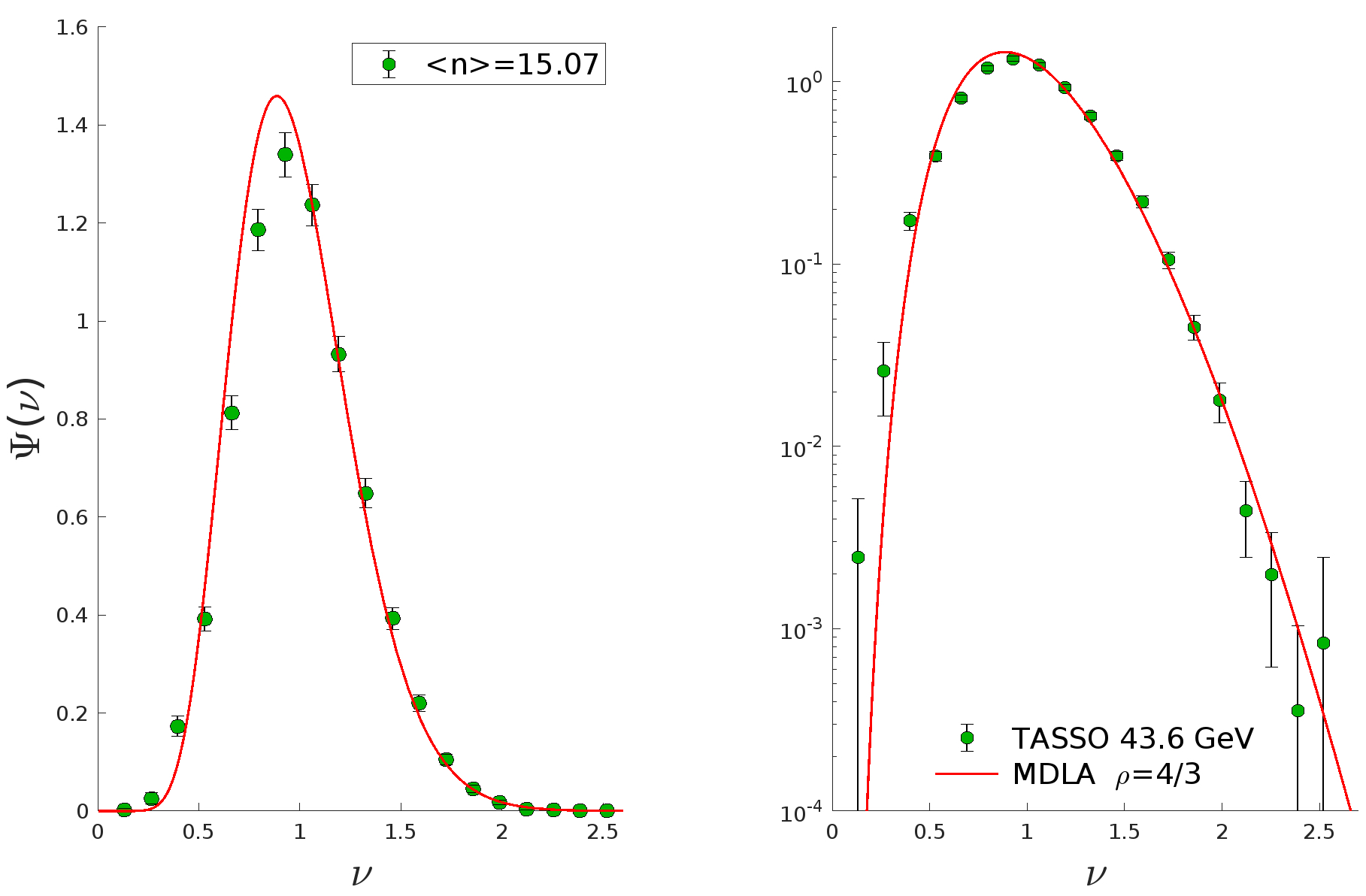}
    \captionof{figure}{P-KNO distribution at $\sqrt{s}=44\,\GeV$; TASSO \cite{TASSO}     \label{fig:TASSO_44}  }
\end{center}

\begin{center}
    \includegraphics[width=0.8\linewidth]{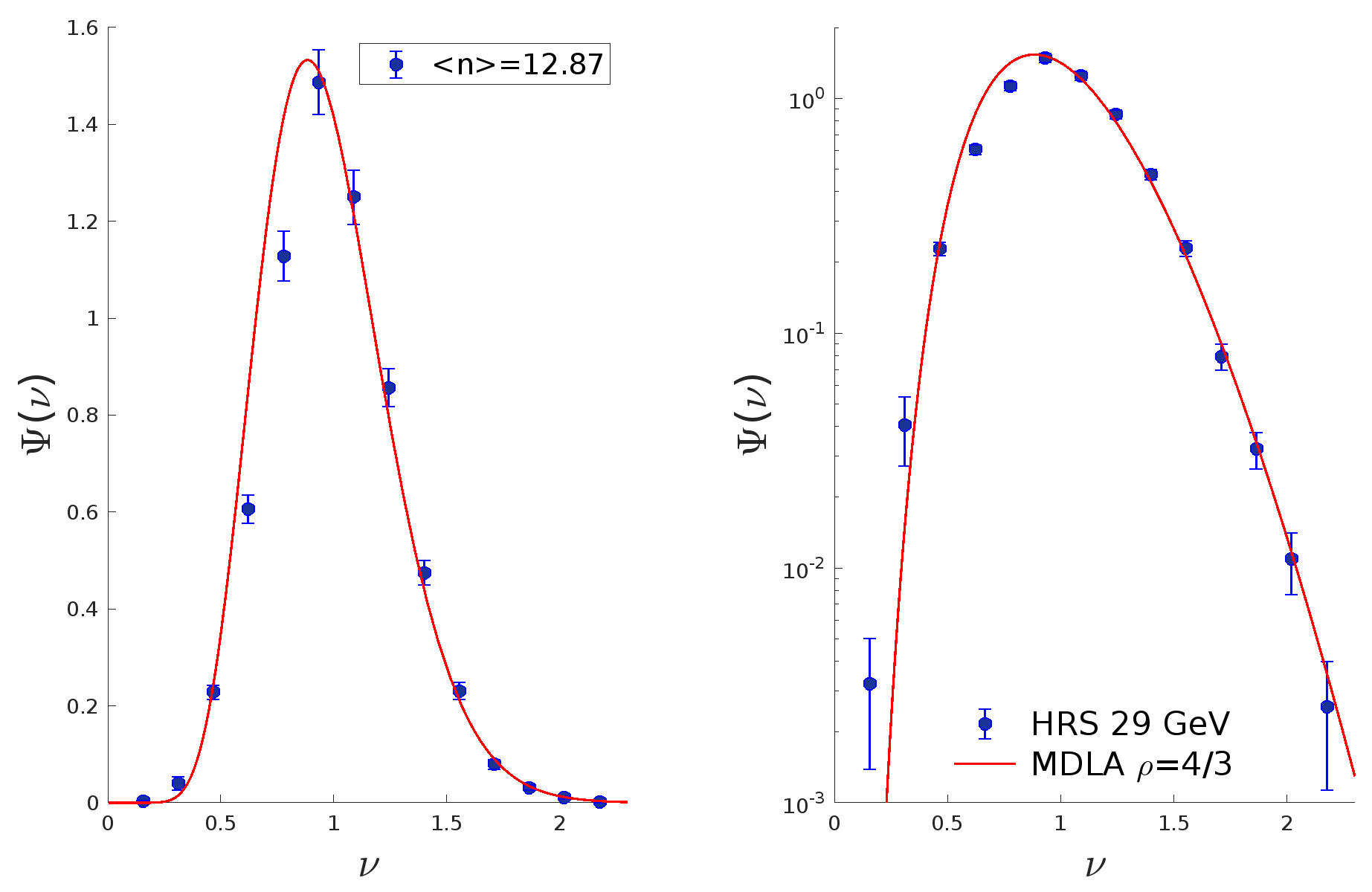}
    \captionof{figure}{P-KNO distribution at $\sqrt{s}=29\,\GeV$; HRS \cite{HRSmoms}     \label{fig:HRS_29}  }
\end{center}

At lower energies, the distributions obtained by ARGUS (10 \GeV) \cite{ARGUS} and TASSO at 14 and 22 \GeV\ are significantly broader than the corresponding MDLA curves. As discussed in Sect.~\ref{Sec:LowMoms}, this could be due to the proximity of the MDLA critical point $\gamma\simeq 0.56$ at low scales.

\subsubsection{Scaling violation}

QCD predicts violation of scaling, a clear sign of which can be observed from comparison of the two panels of Fig. \ref{fig:MOMS_L3_HRS} for the multiplicity moments. 
At a smaller hardness scale (HRS; right panel) the coupling is larger, and so is the multiplicity anomalous dimension. 
This drives down the MDLA factor \eqref{eq:gkgmdla} and therefore the moments with $k\ge 2$.
Smaller values of the moments imply a narrower distribution.

Since the QCD coupling changes slowly, in order to see the effect of narrowing of the P-KNO distribution, precise data and a lever arm in the hardness scale are necessary.  The combination of HRS and LEP-1 measurements provides such an opportunity. 
Violation of scaling is clearly seen in the large-$\nu$ tail of the P-KNO distributions in Fig. \ref{fig:scaling}. 

\begin{center}
    \includegraphics[width=0.8\linewidth]{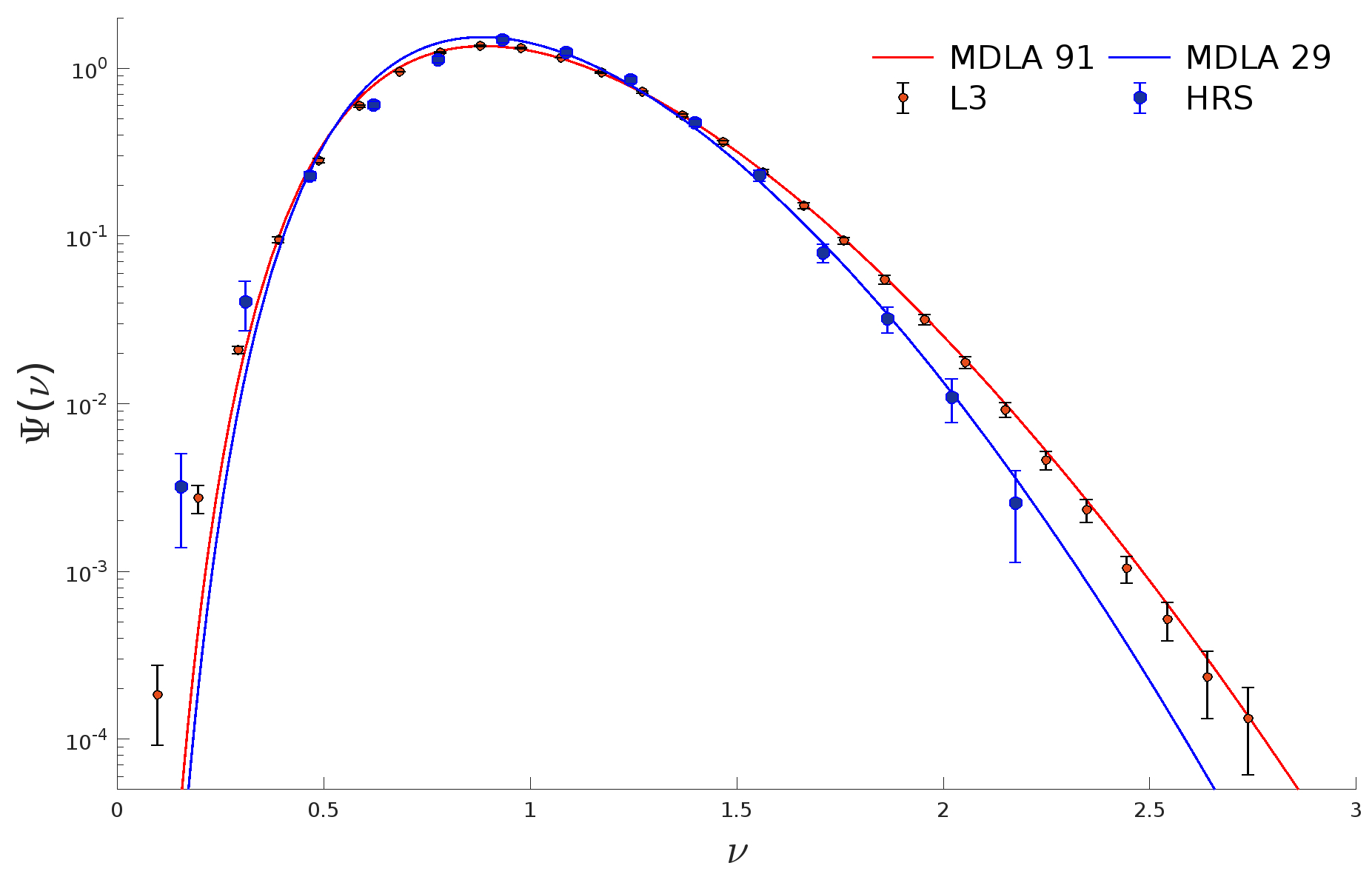}
    \captionof{figure}{P-KNO scaling violation      \label{fig:scaling} }
\end{center}

\subsection{Quark jet}
Most of \ee\ annihilation events have a two-jet structure. By dividing the event into two hemispheres under the guidance of the jet axis, fluctuations in a single-quark (antiquark) jet can be accessed by measuring hadrons in one hemisphere. 

Fig.~\ref{fig:DELPHI_1} shows the P-KNO distribution in one hemisphere measured by DELPHI at the $Z^0$ peak, and Fig.~\ref{fig:HRS_1} by the SLAC HRS Collaboration at $\sqrt{s}=29\,\GeV$. In both cases there is a good agreement with the MDLA predictions. The solid and dashed curves show the predictions for hardness scales $Q=\sqrt s$ and $\sqrt{s/2}$, respectively. As discussed in Sect.~\ref{Sec:PKNO}, the latter may be more appropriate for a single hemisphere, but in practice the difference is within the experimental and theoretical uncertainties.

\begin{center}
    \includegraphics[width=0.8\textwidth]{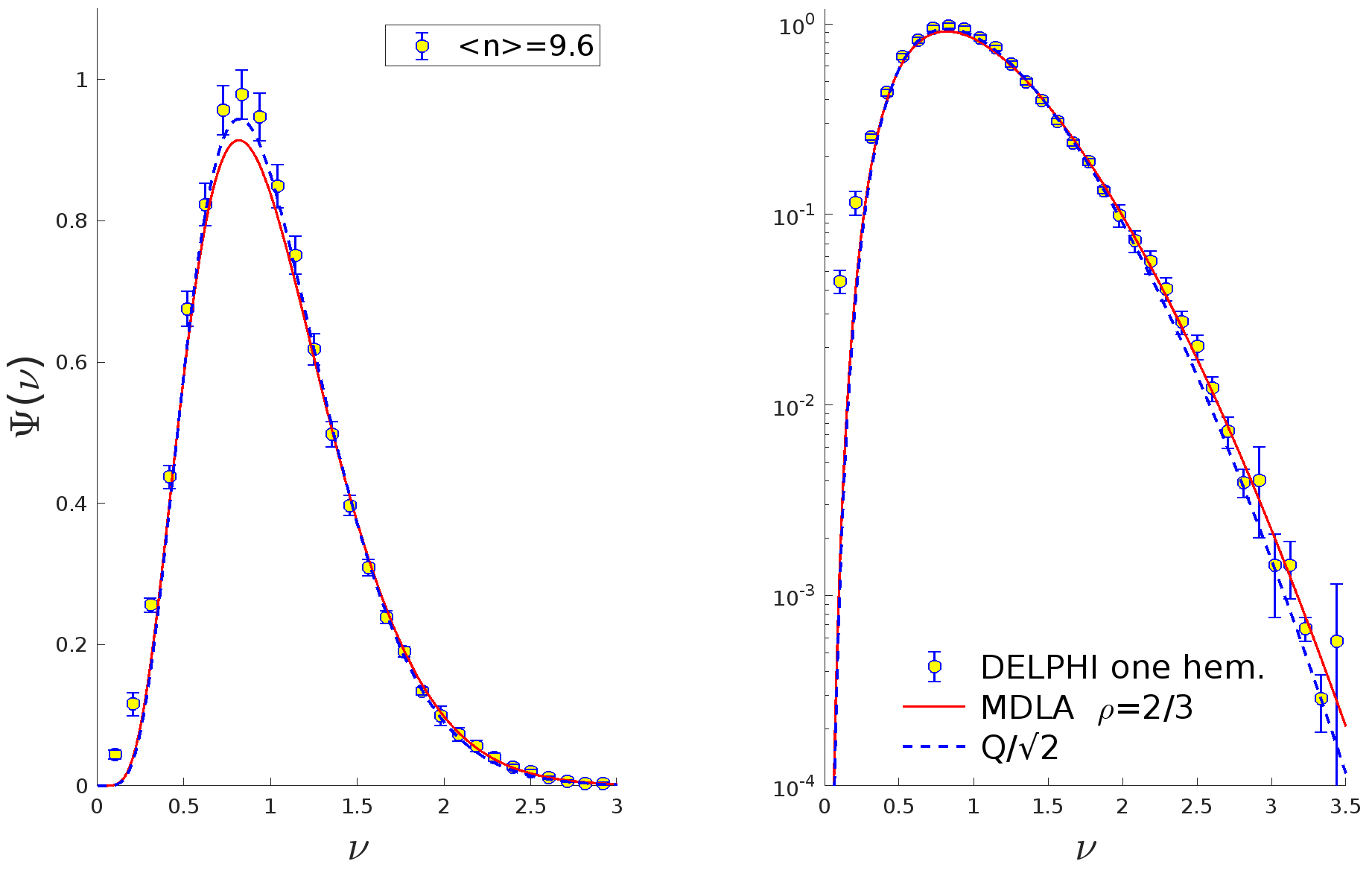}
    \captionof{figure}{Multiplicity fluctuations in one hemisphere; $Q=91.2\,\GeV$, DELPHI \cite{DELPHI1}     \label{fig:DELPHI_1}  }
\end{center}

\begin{center}
    \includegraphics[width=0.8\textwidth]{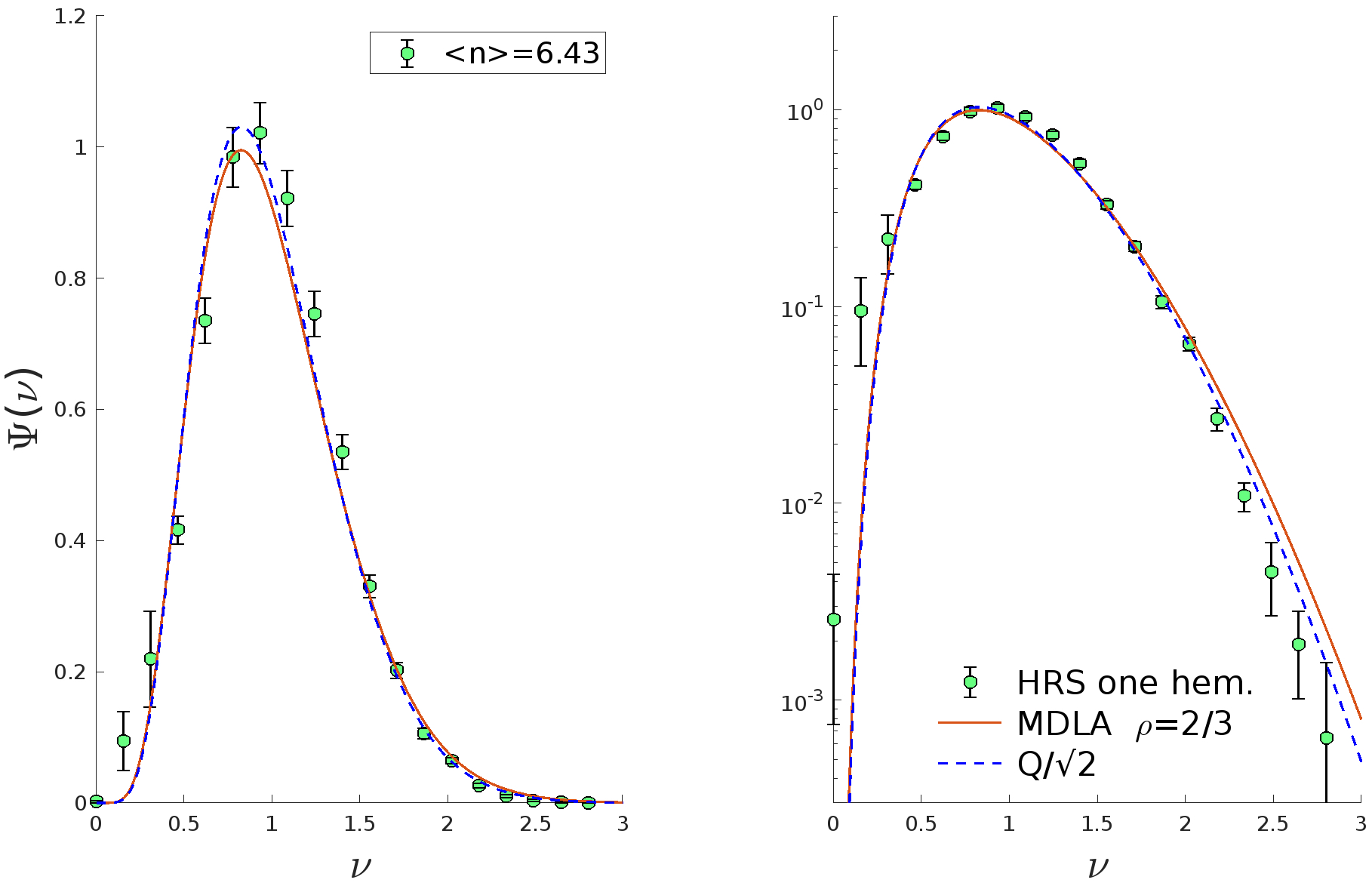}
     %   \vspace{-2mm}
    \captionof{figure}{Multiplicity fluctuations in one hemisphere; $Q=29\,\GeV$, HRS \cite{HRSmoms}     \label{fig:HRS_1}  }
\end{center}

\subsection{Gluon jet}\label{Sec:glujet}

There is a limited amount of data on hadron multiplicity fluctuations in gluon jets.

\noindent
\begin{minipage}{0.4\textwidth}
The OPAL Collaboration measured charged-hadron multiplicity distributions in unbiased gluon jets with energies from 5.3 to 17.7 \GeV\ that were extracted from three-jet \ee\ annihilation events at the $Z$ peak \cite{OPALgE}.  % {OPAL2002}.
In Fig.~\ref{fig:OPAL_g_14} one of these distributions is shown at an intermediate jet energy $E_g^*\simeq 14$, together with the MDLA prediction for $\rho=1$ and the hardness scale $Q=2E_g^*=28\,\GeV$.

For the sake of comparison, the quark multiplicity distribution at the same hardness scale is plotted by the dashed line.  The gluon distribution is somewhat favoured by the data, especially in the tail $\nu\!>\!1$, where the prediction is not model-dependent.
\end{minipage}
\begin{minipage}{0.6\textwidth}
\begin{center}
\includegraphics[width=\textwidth]{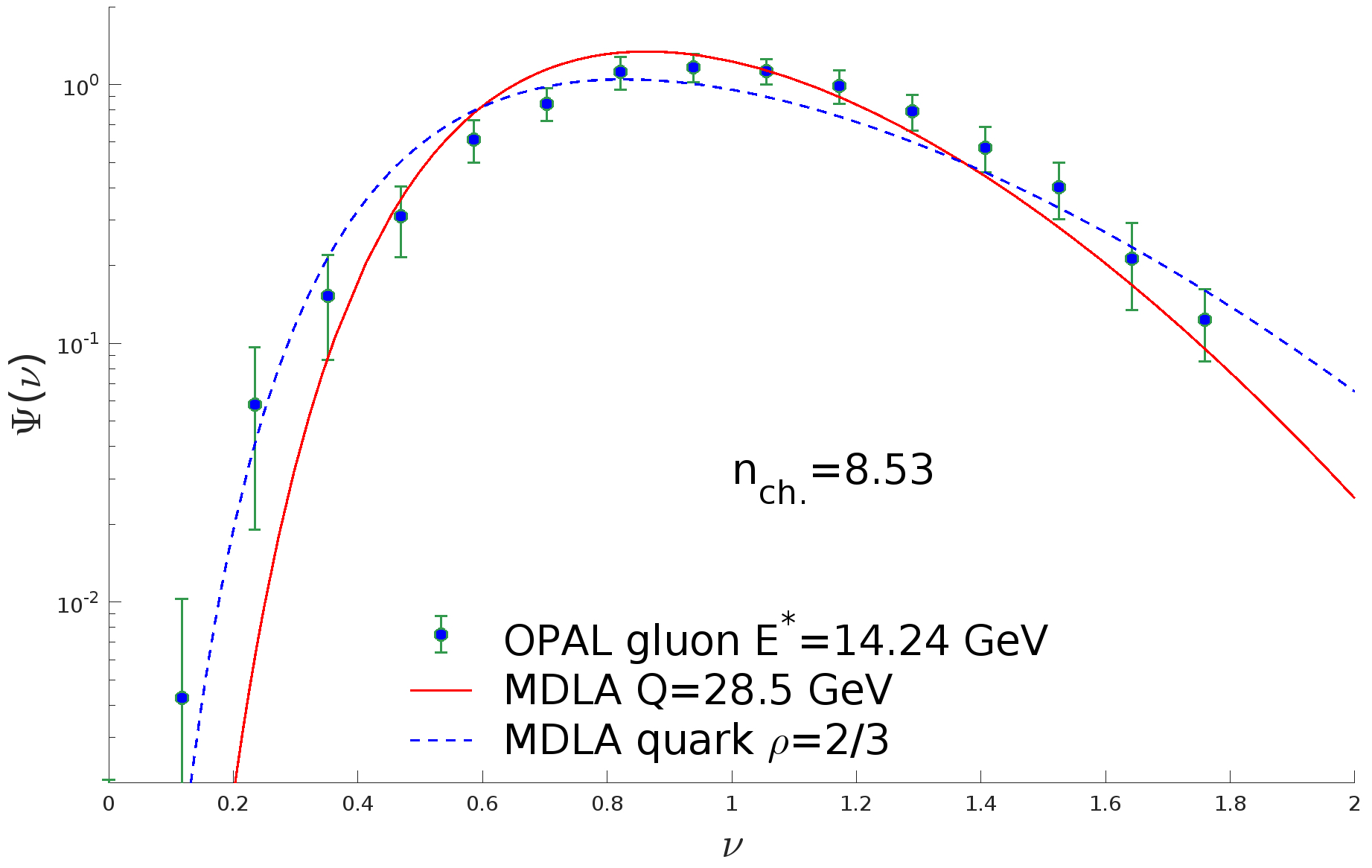}
\captionof{figure}{\label{fig:OPAL_g_14} OPAL gluon jet \cite{OPALgE} vs. MDLA $\rho=1$ }
\end{center}
\end{minipage}

OPAL also studied gluon jets with mean energy 40.1 \GeV\ that recoiled against a pair of quarks in the opposite hemisphere.  The mean multiplicity, multiplicity moments, and distributions of charged hadrons within small rapidity intervals were reported.
As an example, the multiplicity distribution of hadrons with rapidity $y<2$ relative to the gluon jet axis is plotted in Fig. \ref{fig:OPAL_g_80.png} and shows reasonable agreement with the MDLA curve.
\begin{center}
\includegraphics[width=0.8\textwidth]{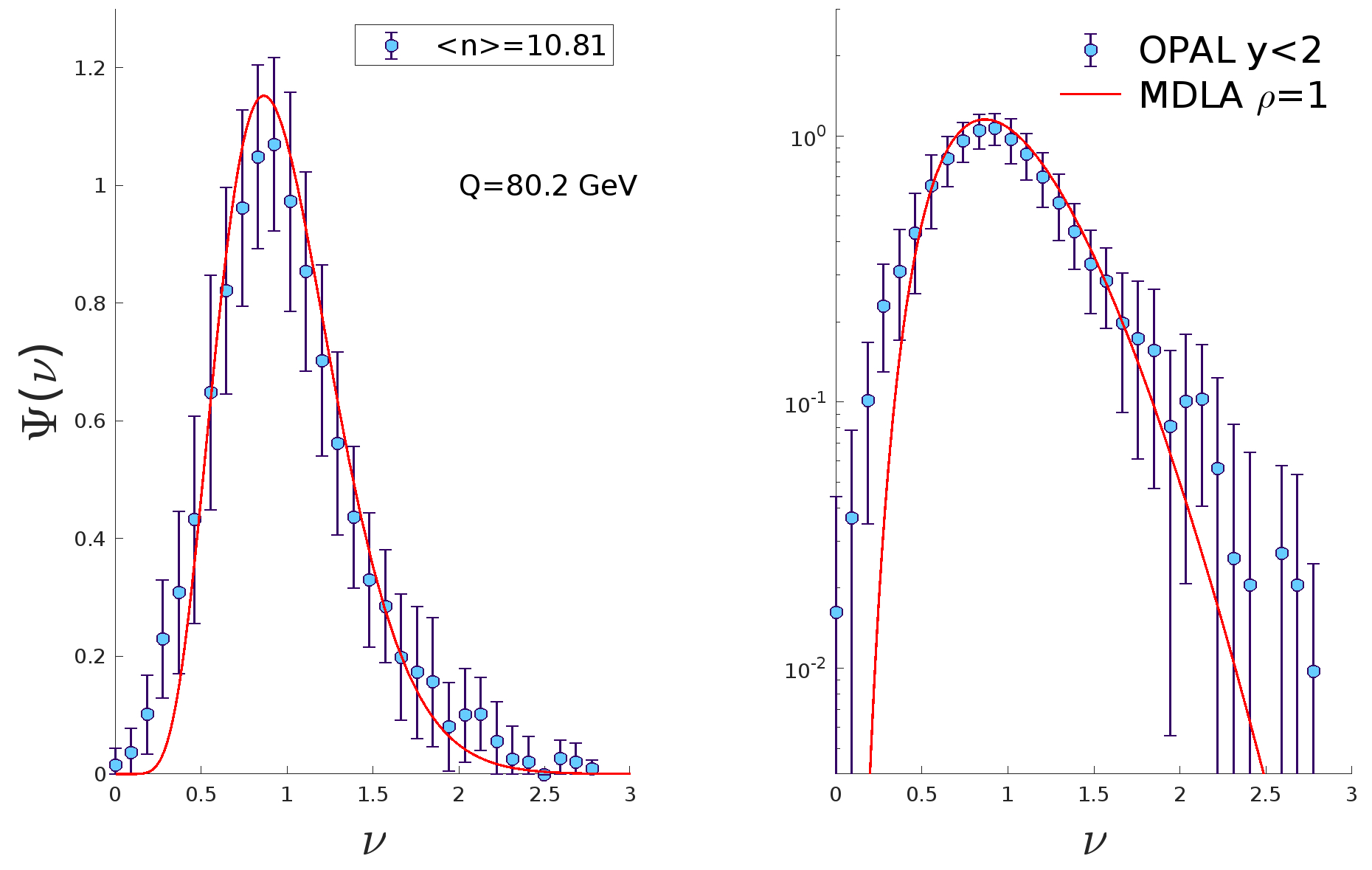}
\captionof{figure}{\label{fig:OPAL_g_80.png} Charged hadron production at small (pseudo)rapidity w.r.t.\ gluon jet axis, OPAL \cite{OPALgqpoint}}
\end{center}
Measuring hadrons in restricted rapidity intervals \citm{OPALgqpoint}{DELPHI-Y} has its peculiarities, a discussion of which we postpone to future work.

\mysection{Conclusions}
In Ref.~\cite{DW25} we showed that a parameter-free QCD-based analysis is able to explain the high-multiplicity fluctuations in hadron production in hard processes.  In the present paper we have extended that analysis to the full range of the multiplicity distribution and its associated moments.  The key component of our approach is the modified double-logarithmic approximation (MDLA)~\cite{D93}, which resums terms enhanced by powers of the moment rank of the Polyakov-KNO distribution.  We observe that this resummation is sufficient to bring the resulting moments into surprisingly good agreement with data from \ee\ annihilation over a wide range of energies, even for the very lowest-order moments that determine the overall shape of the distribution.

This observation encouraged us to follow the programme of~\cite{DW25} to derive the approximate form of the entire P-KNO distribution from its moments. Unmodified, the techniques developed there lead to problems of non-positivity in the low-multiplicity region. We propose a simple two-parameter ansatz for subleading terms that cures these problems, while having negligible effects on the moments and the high-multiplicity region at LEP energies. Furthermore, an energy- and process-independent choice of parameters substantially improves the agreement with low moments at lower energies. Apart from this choice, the predicted distribution depends on a single dynamical quantity, the multiplicity anomalous dimension $\gamma(Q)$, given in terms of the QCD running coupling $\as(Q)$, and, in the case of non-gluon sources, an empirical measure $\rho$ of hadron production power.

The resulting form of the P-KNO distribution is found to be in reasonable quantitative agreement with data from \ee\ annihilation, on the full event multiplicity as well as that of single quark and gluon jets, over a wide range of energies. For the first time, clear evidence is seen for the KNO scaling violation predicted by QCD.  

We end with a few words of caution and possible directions for further work.
\ben

\item 
It remains true that only the high-multiplicity side of the P-KNO distributions is under perturbative QCD control. A theoretical approach to low-multiplicity fluctuations is lacking and should be developed.  In our model, there is a tendency to undershoot the data at the very lowest multiplicities and energies. This might or might not be alleviated by parameter tuning, but more theoretical understanding would be preferable.

\item 
Monte Carlo models based on parton showers are doing well in describing hadroproduction in hard processes. However, on the theory side, a better understanding, and a quantitative model, of non-perturbative effects in the hadronisation of partons is needed.  In particular, the quantity we have called $\rho$, the hadron production power, needs to be clarified.

\item
Apart from the OPAL results discussed in Sect.~\ref{Sec:glujet}, we could not find any data on multiplicity distributions in gluon jets from the other LEP Collaborations. That is a pity. If the data are still accessible, we encourage them to test our predictions more rigorously. Fig.~\ref{fig:OPAL_g_14} suggests that with more statistics the expected difference from quark jets could be established (or otherwise).

\item
Data on hadron multiplicity fluctuations in restricted rapidity intervals, relative to the jet axis, exhibit a strong dependence on the interval selected.  An in-depth study of this phenomenon is warranted.

\item 
The approach developed here can --- and should --- be applied to LHC jets.  For this purpose, a clear separation of data on high-$p_t$ quark and gluon jets would be most helpful. 
\een
\newpage

\par \vskip 1ex
\noindent{\large\bf References}
\begin{enumerate}

\item\label{DW25}
Yu.L.  Dokshitzer and B.R. Webber, \\
``Hadron multiplicity fluctuations in perturbative QCD'',
[arXiv:2505.00652 [hep-ph]], \\
to appear in JHEP.

\item\label{D93}
Yu.L.  Dokshitzer,  
 ``Improved QCD treatment of the KNO phenomenon'',  \\
 \pl{305 }{295}{1993}.   
 % 295-301

\item\label{KNO}
Z.  Koba,  H.B.  Nielsen and P.  Olesen, 
``{Scaling of multiplicity distributions in high-energy hadron collisions}", 
\np{40}{317}{1972}.

\item\label{AMPolyakov} 
A.M.  Polyakov,  
``{A Similarity hypothesis in the strong interactions. 1. Multiple hadron production in \ee\ annihilation}", 
\spj{32}{296}{1971};  \\
``{Similarity hypothesis in strong interactions. 2. Cascade formation of hadrons and their energy distribution in \ee\ annihilation}",
\ib{33}{850}{1971}.  

\item\label{BCM}       
 A. Bassetto,  M. Ciafaloni and  G. Marchesini,    
``{Jet Structure and Infrared Sensitive Quantities in Perturbative QCD}", 
   \prep{100}{201}{1983}.
%    pages = "201--272",
   
 \item\label{DFK}         
   Yu.L.  Dokshitzer,  V.S. Fadin and V.A.  Khoze, 
 ``{Double Logs of Perturbative QCD for Parton Jets and Soft Hadron Spectra}", 
    \zp{15}{325}{1982}.  
    
\item\label{Book} 
       Yu.L. Dokshitzer,  V.A. Khoze,  A.H. Mueller and S.I. Troyan, 
       {\em Basics of Perturbative QCD}, ed. J.\ Tran Thanh Van, 
       (Editions Fronti{\`e}res, 1991).

% https://www.lpthe.jussieu.fr/~yuri/BPQCD/BPQCD.pdf

\item\label{MW84}
E.D.  Malaza and B.R.  Webber,  
 ``QCD Corrections to Jet Multiplicity Moments", 
\pl{149}{501}{1984}.

\item\label{ALEPHdist}
D.~Decamp \textit{et al.} [ALEPH],
``Measurement of the charged particle multiplicity distribution in hadronic Z decays'',
\pl{273}{181}{1991}.

 \item\label{OPAL1992} 
 P.D. Acton  \textit{et al.} [OPAL], 
 ``{A Study of charged particle multiplicities in hadronic decays of the $Z^0$}",
  \zp{53}{539}{1992} .

\item\label{DELPHI1} 
P. Abreu \textit{et al.} [DELPHI],
"{Charged particle multiplicity distributions in $Z^0$ hadronic decays}",
\zp{50}{185}{1991}.

\item\label{L3} 
B. Adeva \textit{et al.} [L3],
``Studies of hadronic event structure and comparisons with QCD models at the $Z^0$ resonance'',
\zp{55}{39}{1992}.

\item\label{HRSmoms}
    M. Derrick \textit{et al.} [HRS],
``{Study of Quark Fragmentation in \ee\ Annihilation at 29 GeV: Charged Particle Multiplicity and Single Particle Rapidity Distributions}",
 \pr{34}{3304}{1986}.

\item\label{AHM}
A.H. Mueller, 
``Multiplicity and Hadron Distributions in QCD Jets: Nonleading Terms", \\
\np{213}{85}{1983}.

\item\label{BW84}    
 B.R. Webber,    ``Average Multiplicities in Jets", 
   \pl{143}{501}{1984}.

\item\label{EAO}
Yu.L. Dokshitzer  and S.I. Troyan,  
``Asymptotic Freedom and Local Parton--Hadron Duality", 
Proceedings of the XIX Winter School of the LNPI,  vol. 1,  page 144.  Leningrad, 1984.

\item\label{OPAL_LEP2} 
K. Ackerstaff \textit{et al.} [OPAL], 
"{QCD studies with \ee\ annihilation data at 161-GeV}",
\zp{75}{193}{1997}; \\
G. Abbiendi \textit{et al.} [OPAL],
``QCD studies with \ee\ annihilation data at 172 GeV - 189 GeV",
  %  eprint = "hep-ex/0002012",
  \epj{16}{185}{2000}.

 \item\label{TASSO} 
W. Braunschweg  \textit{et al.} [TASSO], 
``Charged Multiplicity Distributions and Correlations in \ee\ Annihilation at PETRA Energies", 
\zp{45}{193}{1989}.

\item\label{ARGUS} 
H. Albrecht \textit{et al.} [ARGUS],
"{Measurement of R and determination of the charged particle multiplicity in \ee\ annihilation at $\sqrt s$ around 10-GeV}",
  \zp{54}{13}{1992}.

\item\label{OPALgE}
% gluon jets 5-18 GeV
G.~Abbiendi \textit{et al.} [OPAL],
``Experimental studies of unbiased gluon jets from \ee\ annihilations using the jet boost algorithm''
\pr{69}{032002}{2004}.

\item\label{OPALgqpoint}
% gluon recoiling against qq.  40.1 GeV point
G. Abbiendi \textit{et al.} [OPAL], 
``Experimental properties of gluon and quark jets from a point source'', 
\epj{11}{217}{1999}.
% e-Print: hep-ex/9903027 [hep-ex]

\item\label{ALEPH-Y}
D. Buskulic \textit{et al.} [ALEPH],
``Measurements of the charged particle multiplicity distribution in restricted rapidity intervals'',
\zp{69}{15}{1995}.

\item\label{DELPHI-Y} 
P. Abreu \textit{et al.} [DELPHI],
"{Charged particle multiplicity distributions in restricted rapidity intervals in Z0 hadronic decays}",
\zp{52}{271}{1991}.

\item\label{CMW}
 S. Catani,  G. Marchesini and B.R. Webber,  
 ``QCD coherent branching and semiinclusive processes at large $x$", 
 \np{349}{635}{1991}.

\end{enumerate}

\appendix

\section*{Appendices \label{App:A1}}

\mysection{Running QCD coupling $\as$  \label{App:as}}

The 2-loop coupling satisfies the differential equation 
\begin{subequations}\label{eq:2loop}
\beq
 \frac{d}{d\ln Q} \left( \frac{2\pi}{\as(Q^2)}\right) =  \beta_0 + \frac12\beta_1  \left(\frac{\as(Q)}{2\pi}\right)
\eeq
with
\beq
 \beta_0 = \frac{11}3N_c-\frac23 n_f = 11 - \frac23 n_f  ,  \quad \beta_1 = \frac{34}3N_c^2 -2C_F n_f - \frac{10}3N_c  n_f =  {102} -  \frac{38}{3}n_f .
\eeq
\end{subequations}
In \cite{DW25} we used the approximate solution with $Y=\ln(Q/\LQCD)$ kept in the denominator:
\beq\label{eq:Aden}
  \frac{\as(Q)}{2\pi} = \left( \beta_0 Y + \frac{\beta_1}{2} \ln Y + \mbox{const} \right)^{-1}
  = \left( \beta_0 Y + \frac{\beta_1}{2} \ln \frac{Y}{\ln(M_Z/\LQCD)} \right)^{-1}.
\eeq
With the integration constant fixed in this way, the QCD scale parameter is expressed through the coupling value at $Q\!=\!M_Z$ in a simple way:
\beq
  \LQCD = M_Z \exp \left\{ - \frac{2\pi}{\beta_0\as(M_Z)} \right\}.
\eeq

In the present work, however, we employ the more common expanded version of \eqref{eq:Aden},
\beq\label{eq:Aexp}
  \frac{\as(Q)}{2\pi} =  \frac{1}{\beta_0 Y} - \frac{\beta_1 }{2\beta_0^2 \,Y^2}\ln\frac{\beta_0 \,Y }{2\pi \as(M_Z)}\>+\> \cO{\as^3}.
\eeq
The difference between the two prescriptions \eqref{eq:Aden} and \eqref{eq:Aexp} is hardly visible above $Q=2\,\GeV$.
\begin{center}
    \includegraphics[width=0.8\linewidth]{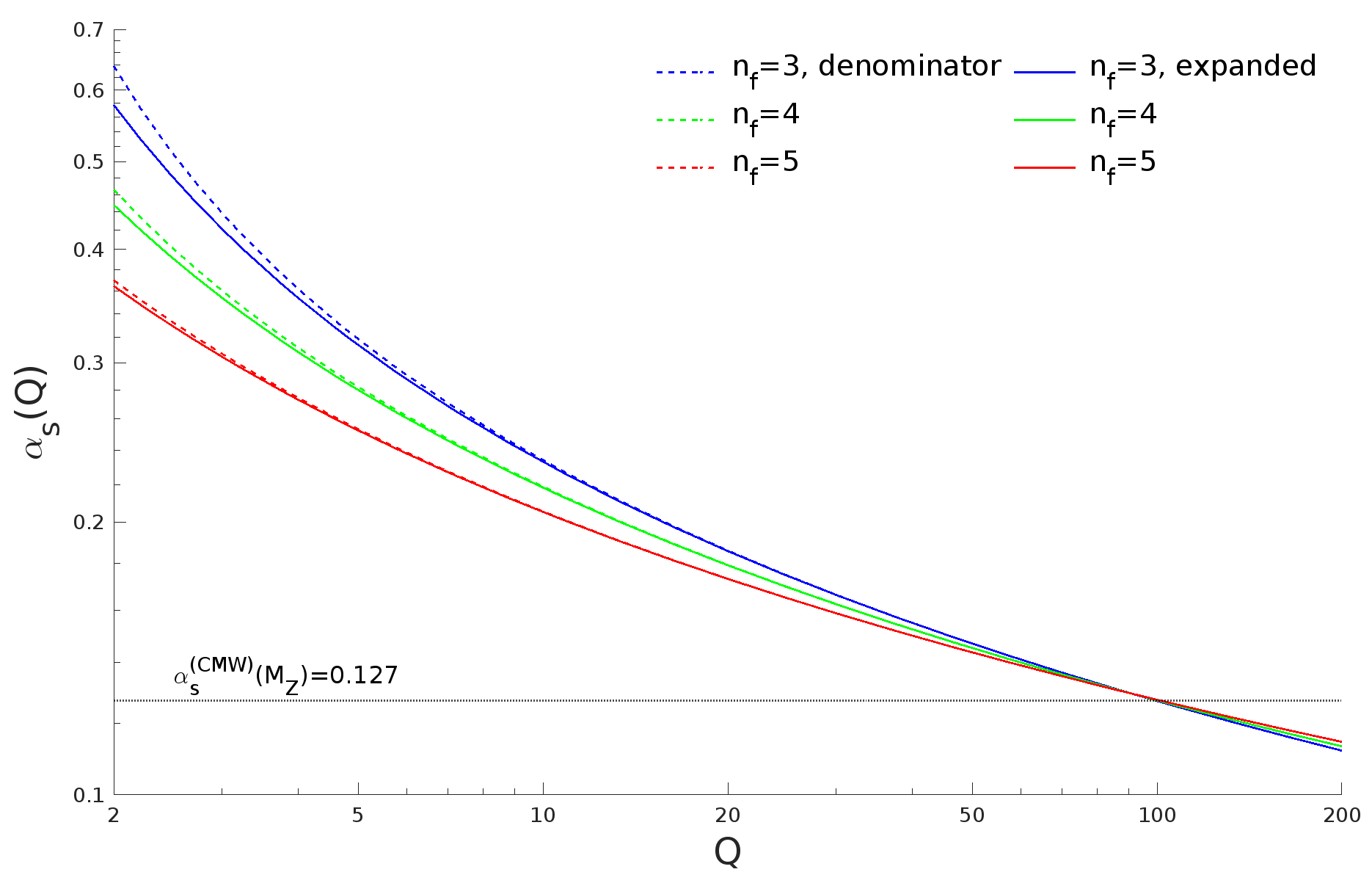}
        \captionof{figure}{Two-loop running QCD coupling     \label{fig:A_as} }
\end{center}

We remind the reader that since multiplicities are driven by soft-gluon radiation, we employ the {\em physical scheme }\/ for the coupling, also known as the ``bremsstrahlung" or CMW \cite{CMW} scheme.  
This translates the MS-bar coupling $\alpha(M_Z)=0.119 $ into the physical coupling $\as=0.127$ (for 5 quark flavours), yielding $\LQCD=0.412\,\GeV $.

%%%%%%%%%%%%%%%%%%%%%%%%%%%%%%
\mysection{Multiplicity anomalous dimension $\gamma$  \label{App:gam}}

For the anomalous dimension of mean multiplicity we use the two-loop QCD expression \cite{BW84}
\beq\label{eq:2lcoup}
\gamma(\as) \>=\> \sqrt{2N_c\frac{\as}{\pi}} - \left( \frac{\beta_0}{4} + \frac{10n_f}{3N_c^2}\right) \frac{\as}{2\pi}, \quad  \beta_0 = \frac{11}3N_c-\frac23 n_f . %  = 11 - \frac23 n_f .
\eeq

\begin{center}
    \includegraphics[width=0.8\linewidth]{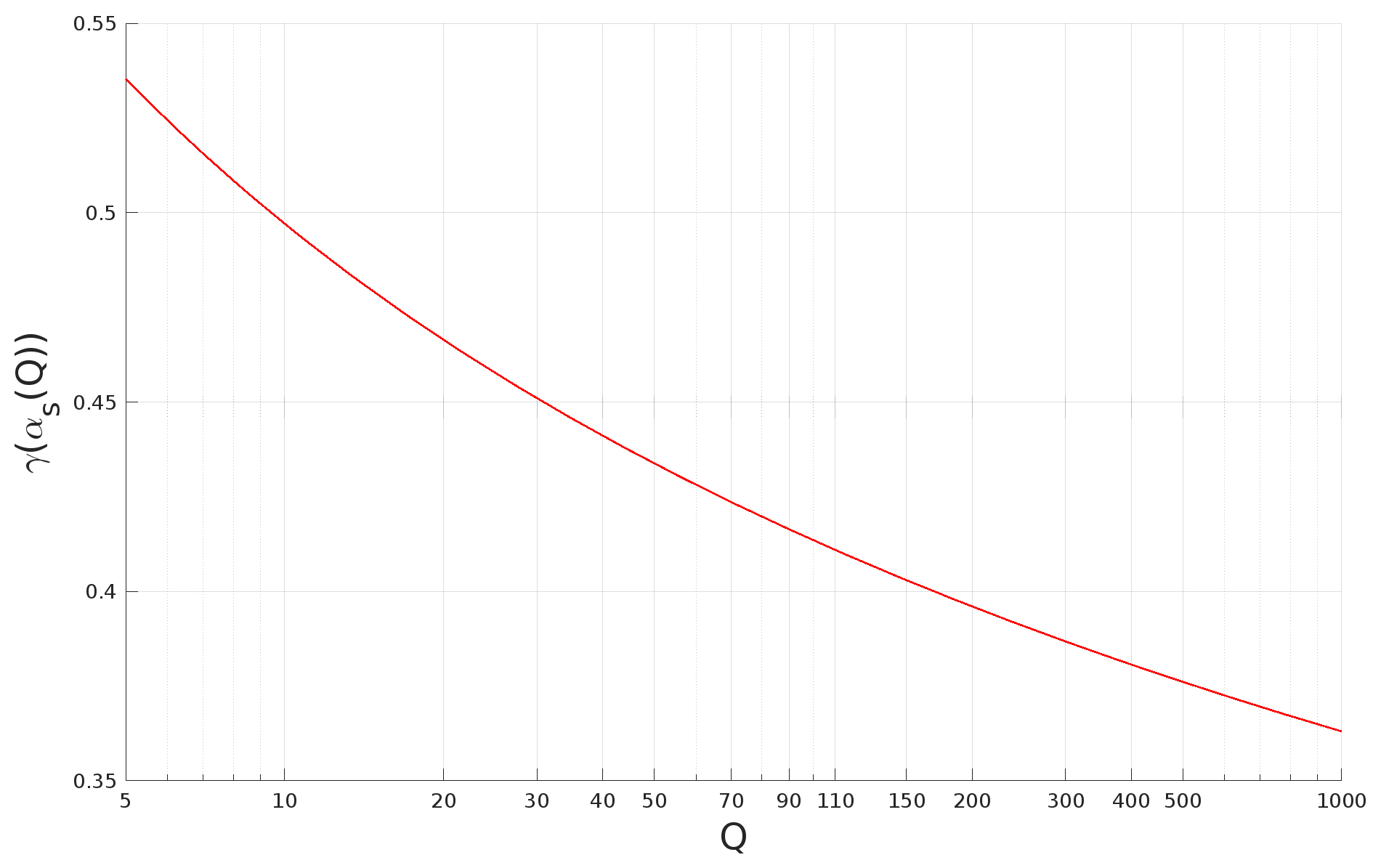}
    \captionof{figure}{Multiplicity anomalous dimension $\gamma(\as(Q))$   \label{fig:A_gam} }
\end{center}

%%%%%%%%%%%%%%%%%%%%%%%%%%%%%%  
\mysection{MDLA factor  \label{App:gkMD}}
The scale-dependent MDLA factor that modifies the asymptotic DLA multiplicity moments in \eqref{eq:gkgmdla} is plotted as a function of the mean multiplicity anomalous dimension $\gamma$ in Fig. \ref{fig:A_gkMD}.
\begin{center}
    \includegraphics[width=0.8\linewidth]{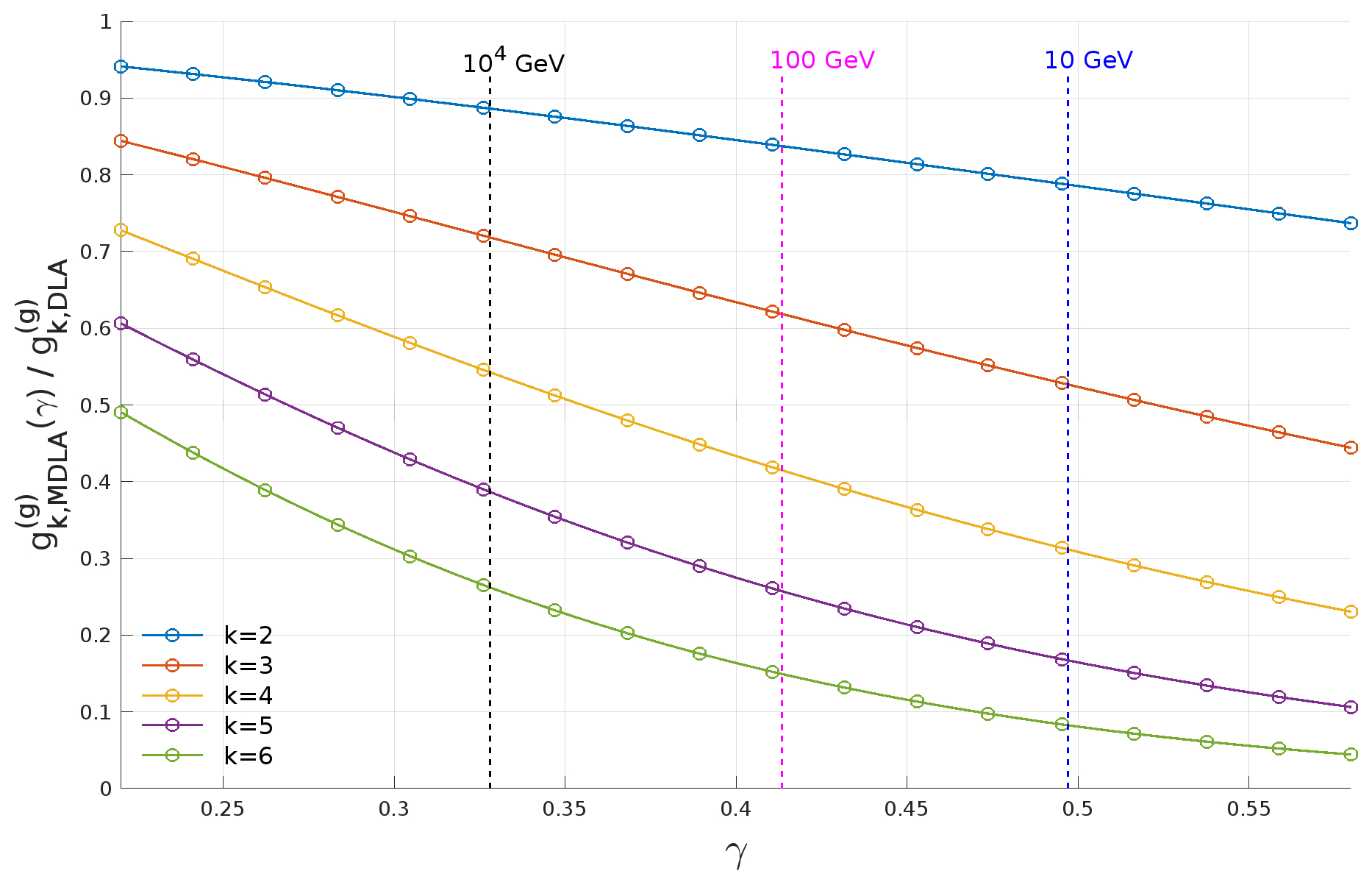}
    \captionof{figure}{The MDLA suppression factor for low multiplicity moments     \label{fig:A_gkMD} }
\end{center}

%%%%%%%%%%%%%%%%%%%%%%%%%%%%%%
\mysection{Polyakov exponent $\mu(\gamma)$  \label{App:mu}}

\begin{center}
    \includegraphics[width=0.8\linewidth]{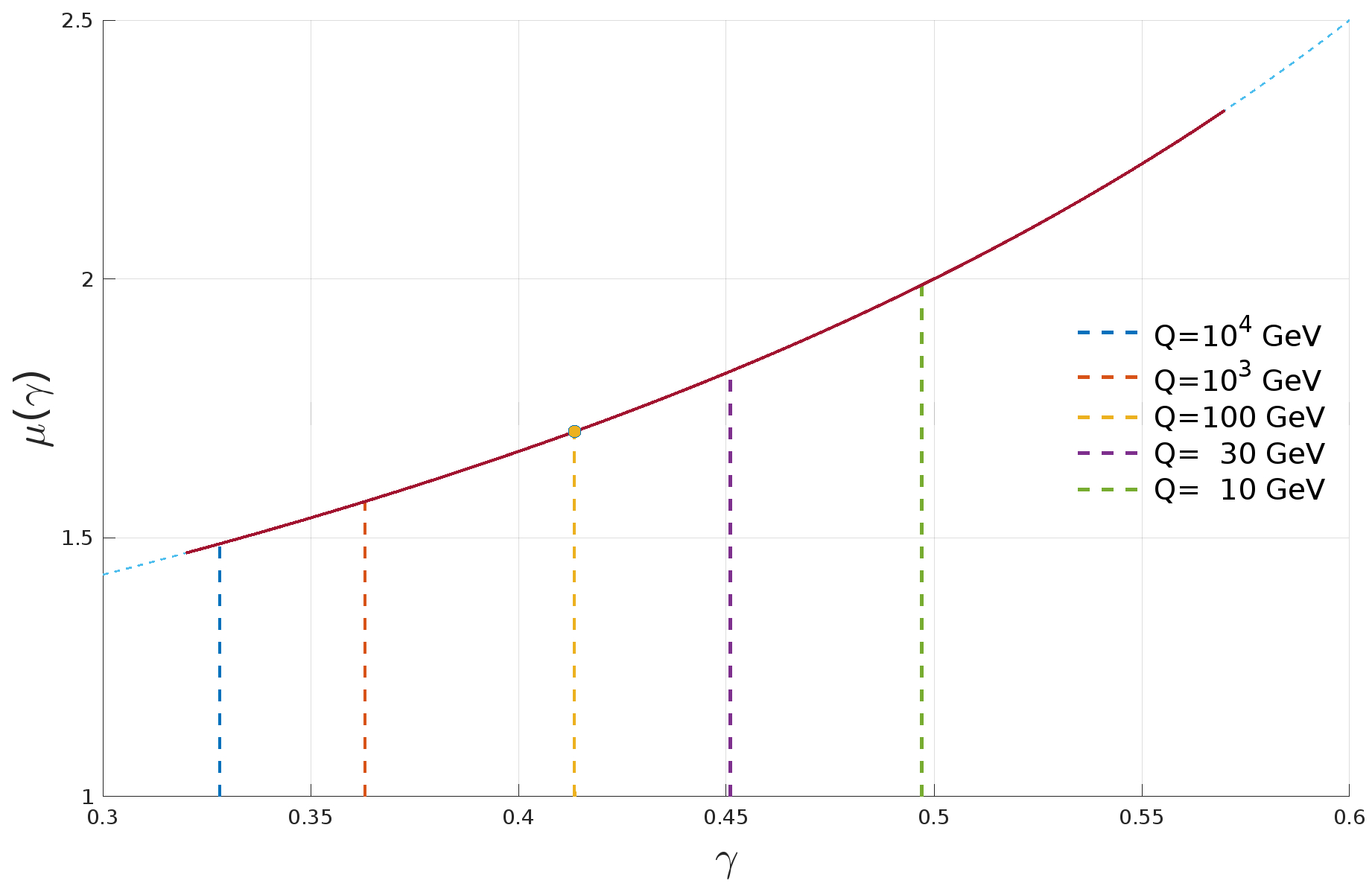}
    \captionof{figure}{Polyakov exponent $\mu(\gamma)$ with characteristic hardness scales marked    \label{fig:A_mu} }
\end{center}

%%%%%%%%%%%%%%%%%%%%%%%%%%%%%%
\mysection{Function $D(\gamma)$ in the Polyakov exponential  \label{App:D}}
The function $D$ multiplies the multiplicity ratio $\nu$ in the Polyakov exponential \eqref{eq:PolExp} 
  \begin{center}
  \includegraphics[width=0.8\linewidth]{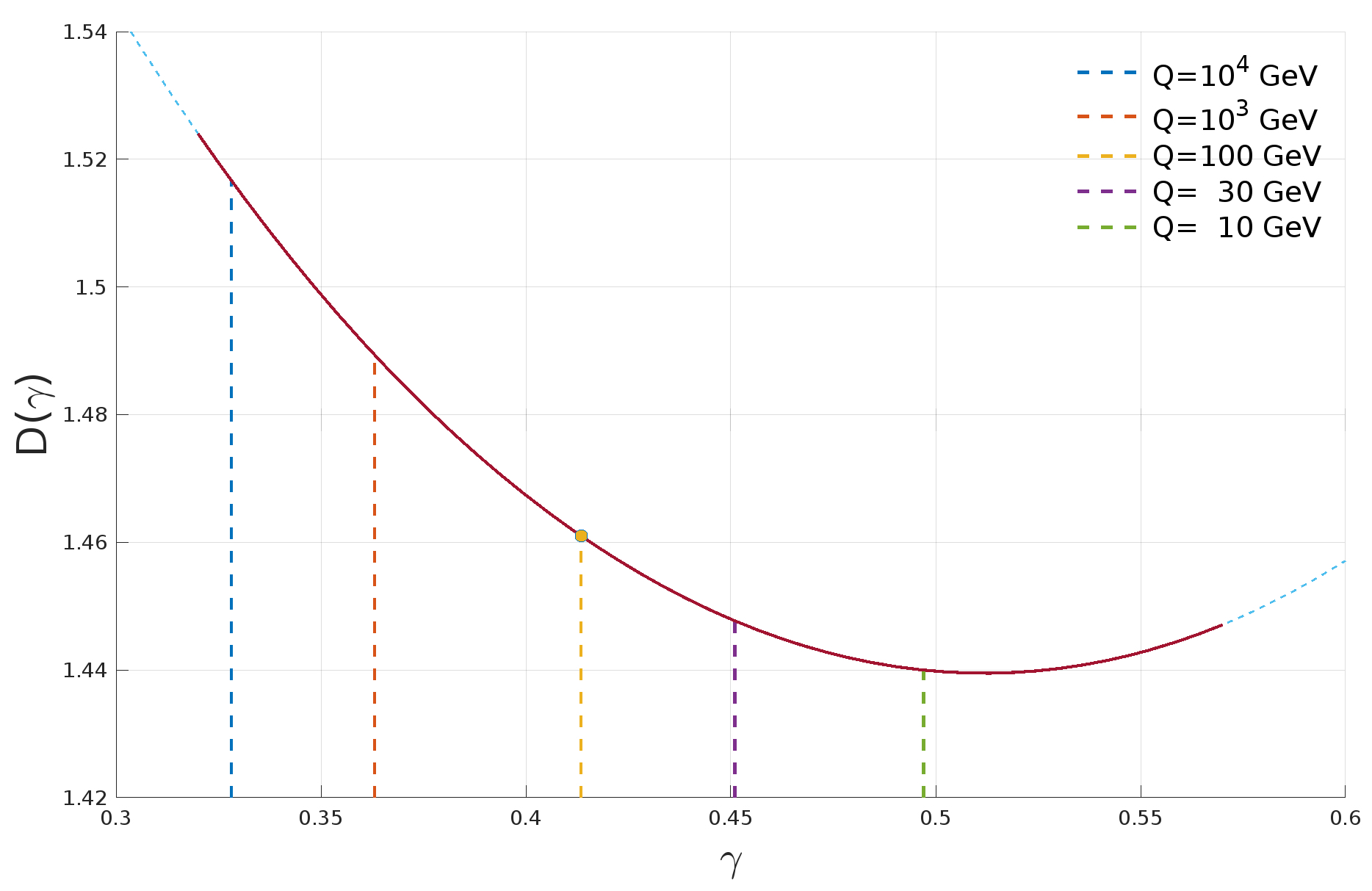}
    \captionof{figure}{The P-KNO exponent multiplier $D(\gamma)$ with characteristic hardness scales marked   \label{fig:A_D}} 
\end{center}

%%%%%%%%%%%%%%%%%%%%%%%%%%%%%%
\mysection{Characteristic moment rank $\ksd$  \label{App:ksd}}
The function $\ksd(\nu, \gamma)$ appears in the master Eq. \eqref{eq:PsiGS} as a result of reconstructing the gluon P-KNO distribution $\Psi(\nu)$ from its Laplace moments. This procedure involves a steepest descent evaluation (therefore, the subscript $\ksd$); see Appendix D of \cite{DW25}.
  \begin{center}
    \includegraphics[width=0.8\linewidth]{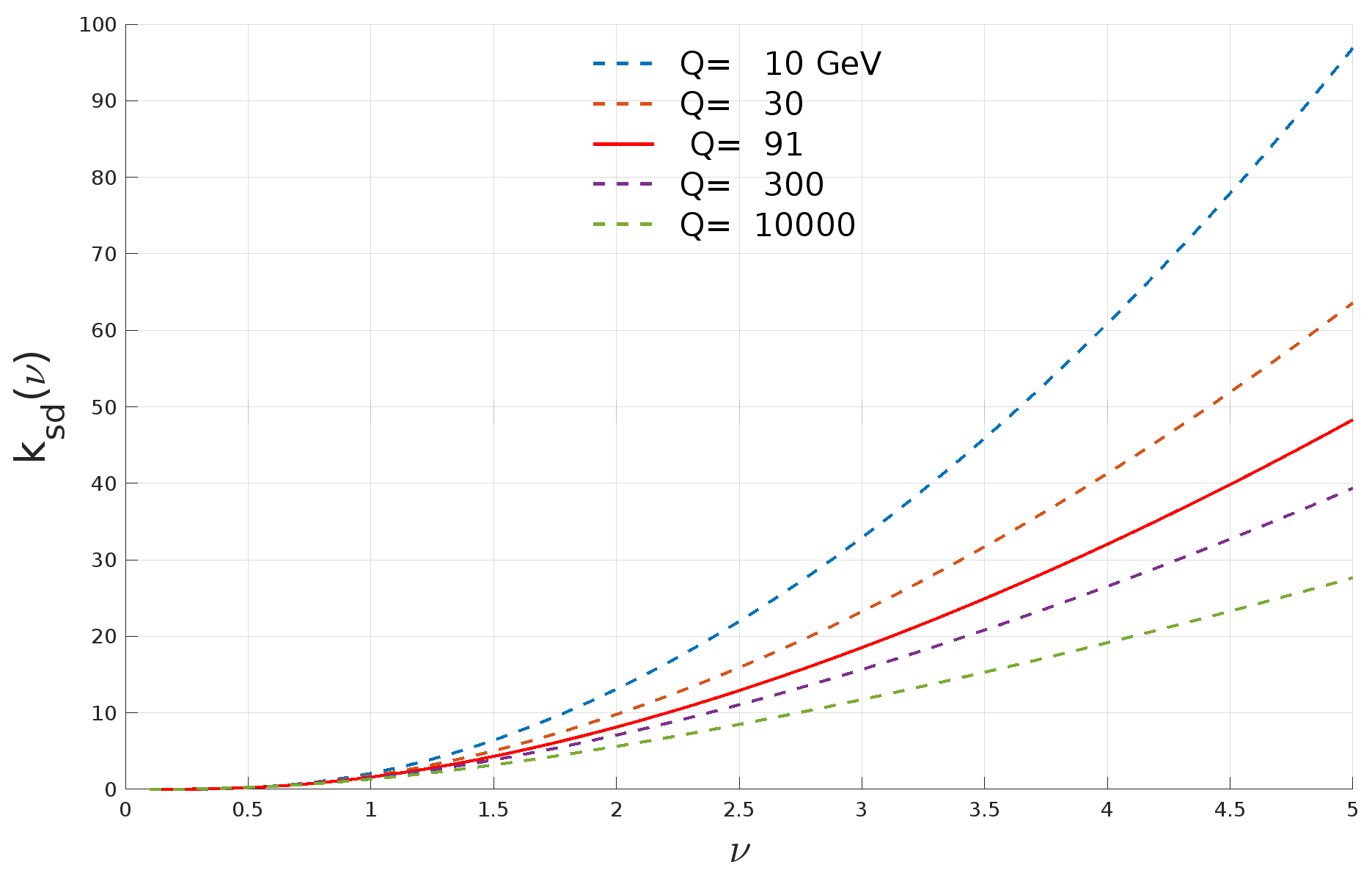}
    \captionof{figure}{Characteristic moment rank $\ksd(\nu, \gamma)$ at different hardness scales     \label{fig:A_ksd} }
  \end{center}
  
%%%%%%%%%%%%%%%%%%%%%%%%%%%%%%
\mysection{Prefactor $\chi$  \label{App:chi}}
%%%  chi
The factor $\chi$ is defined by Eq. \eqref{eq:chidef}.
Constructing the ratio of $\Gamma$ functions (App. C of \cite{DW25}), one obtains
\beeq\label{eq:Achi}
\chi(k,\gamma) &\approx & \frac{e}{\sqrt{2\pi}}\, \sqrt{\frac{1+k}{(1+\gamma k)(1+(1-\gamma)k)}} \cdot \frac{\cR(k)}{\cR(\gamma k) \cR((1-\gamma)k) } , \\
\cR(k) &=& \big(1+ k^{-1}\big)^k \left(1+ \frac1{12(1+k)^2} \right)^{(1+k)} .
\eeeq
There is a mismatch between \eqref{eq:Achi} and the expression that was announced as the answer for $\chi(k,\gamma)$ in Eq. (C.7) of \cite{DW25}, due to the unnecessary simplifications made there. 

We stress that the full expression \eqref{eq:Achi} was used for the calculations throughout both papers.

The function $\chi$ depends on the scale via the anomalous dimension $\gamma(\as(Q))$ both explicitly and implicitly through the characteristic moment rank $\chi =\chi(\ksd(\nu, \gamma), \gamma) $.
  \begin{center}
    \includegraphics[width=0.8\linewidth]{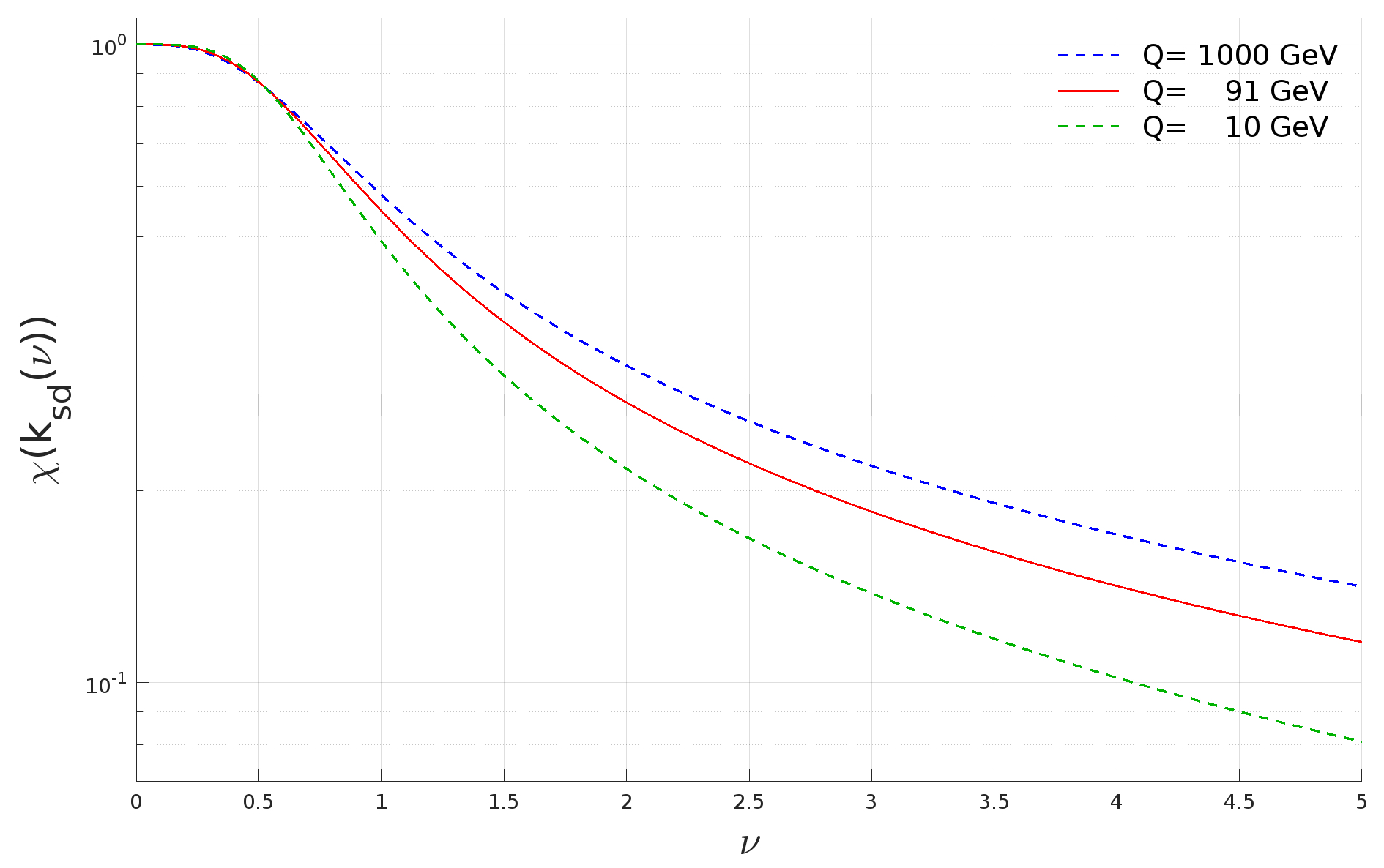}
    \captionof{figure}{The prefactor $\chi(\ksd(\nu,\gamma),\gamma)$ as a function of $\nu$ for different hardness scales     \label{fig:A_chi}}
 \end{center}

\end{document}